\documentclass[nofootinbib,amsmath,amssymb,aps,pre,longbibliography,
  superscriptaddress]{revtex4-2}

\usepackage[T1]{fontenc}
\usepackage[utf8]{inputenc}
\usepackage{graphicx}
\usepackage{hyperref}
\usepackage{tikz}
\usepackage{xcolor}
\usepackage{bm}

\definecolor{darkgreen}{rgb}{0,0.6,0}
\definecolor{darkblue}{rgb}{0,0,0.6}
\definecolor{darkred}{rgb}{0.6,0,0}
\definecolor{darkpurple}{rgb}{0.5,0,0.5}

\hypersetup{
bookmarks=true,
bookmarksopen=true,
pdftitle="",
pdfauthor="", 
pdftoolbar=false, 
pdfstartview={FitH},		
pdfmenubar=true,			
pdfhighlight=/O,			
colorlinks=true,			
urlcolor=darkblue,
citecolor=darkblue,		
linkcolor=darkblue}

\newcommand{\plaind}{\mathop{}\!\mathrm{d}} 

\newcommand{\ve}[1]{ \bm{#1} }

\DeclareMathOperator*{\argmin}{\mathrm{argmin}}
\DeclareMathOperator{\sign}{\mathrm{sign}}

\newcommand{\gammac}{ \gamma_\mathrm{c} }

\newcommand{\A}{ \mathrm{A} }
\newcommand{\B}{ \mathrm{B} }
\newcommand{\C}{ \mathrm{C} }
\newcommand{\D}{ \mathrm{D} }

\newcommand{\sO}{ \mathrm{O} }
\newcommand{\sP}{ \mathrm{P} }
\newcommand{\sQ}{ \mathrm{Q} }
\newcommand{\sR}{ \mathrm{R} }
\newcommand{\sS}{ \mathrm{S} }

\newcommand{\z}{ \ve{z} }

\newcommand{\bigO}{\mathcal{O}}

\newcommand{\basin}{ \mathcal{D} }

\newcommand{\zo}{\varsigma}

\begin{document}

\title{Nonreciprocal dynamics with weak noise: \\aperiodic ``Escher cycles'' and their quasipotential landscape}

\author{Janik Schüttler}
\affiliation{Department of Applied Mathematics and Theoretical Physics, University of Cambridge, Wilberforce Road, Cambridge CB3 0WA, United Kingdom}

\author{Robert L. Jack}
\affiliation{Department of Applied Mathematics and Theoretical Physics, University of Cambridge, Wilberforce Road, Cambridge CB3 0WA, United Kingdom}
\affiliation{Yusuf Hamied Department of Chemistry, University of Cambridge, Lensfield Road, Cambridge CB2 1EW, United Kingdom}

\author{Michael E. Cates}
\affiliation{Department of Applied Mathematics and Theoretical Physics, University of Cambridge, Wilberforce Road, Cambridge CB3 0WA, United Kingdom}

\date{\today}

\begin{abstract}
We present an explicit construction of the Freidlin--Wentzell quasipotential of a stochastic system with two degrees of freedom and nonreciprocal interactions. This model undergoes noise-induced transitions between four metastable attractors, forming recurrent but aperiodic ``Escher cycles,'' similar to the cyclic nucleation dynamics observed in the nonreciprocal Ising model. 
We calculate the quasipotential analytically to first order in nonreciprocality.
We characterise it along a one-dimensional reaction coordinate that connects the attractors, and we also obtain the full two-dimensional landscape, at leading order in perturbation theory. 
The resulting landscapes feature flat regions and extended plateaus, together with non-differentiable switching lines. These singular structures arise from two geometric mechanisms: the handover of dominance between competing transition paths, and the competition between basins of attraction.
The system provides a rare case where the geometry of nonequilibrium rare events can be fully resolved, and a simple analytically tractable example of a quasipotential in more than one coordinate that captures a rich set of nonequilibrium features.
\end{abstract}

\maketitle

\section{Introduction}

This work considers cyclic sequences of transitions between metastable states, reminiscent of motion in a tilted washboard potential; we call these sequences \emph{Escher cycles}. 
They occur in non-equilibrium systems and involve repeated transitions between {distinct} states.
This generic behaviour appears in systems including active flocks, driven spin models, and biological colonies.
 For example, $1d$ active Ising models feature aligned flocks that move coherently before stochastically reversing direction~\cite{oloan1999alternating}.  Similarly, aligned states of incompressible Toner--Tu hydrodynamics are rendered metastable by the nucleation of topological defects \cite{besse2022metastability}, which for finite systems lead to rare (but swift) changes in the direction of long-range orientational order; similar behaviour is also found in the active Ising model {for $d>1$}~\cite{benvegnen2023metastability}.
In motile bacterial colonies undergoing phase separation and logistic growth, colony formation and destruction cycles stochastically between different locations \cite{grafke2017spatiotemporal}. Recent studies of cyclic nucleation in Potts-type models likewise revealed recurrent stochastic switching between ordered states driven by nucleation events~\cite{noguchi2024Spatiotemporal,noguchi2025spatiotemporal}.

{These examples all feature the hallmark of an Escher cycle: long steady-state trajectories display repeated noise-induced transitions among a cycle of metastable states.}
Their name is inspired by M.C. Escher’s artwork \textit{Ascending and Descending}, where figures walk endlessly down staircases that paradoxically loop back to the starting point. Each step of the staircase feels like a descent, but the overall path forms a closed cycle. 
{The idea that descent and ascent might be subjective notions has a natural realisation in systems with non-reciprocal interactions, where different components of the system each act to minimise their own ``selfish energies''~\cite{avni2023non,avni2024dynamical}.  Such non-reciprocal interactions are much-studied in physics, where they are known to produce self-sustained flows, broken time-reversal symmetry, and diverse dynamical phases; they appear in natural systems ranging from active matter, pattern-forming media, and oscillator networks~\cite{you2020nonreciprocity,fruchart2021non,shankar2022topological}.}
{We emphasize that Escher cycles rely on noise to overcome barriers between states, so trajectories are intrinsically aperiodic in time; this may be contrasted with limit cycles, where the system circulates with a well-defined period.}

{In equilibrium systems, rare noise-induced transitions between metastable states often occur by nucleation, and may be described by classical nucleation theory.  Their rates can be estimated by identifying a suitable reaction coordinate and computing the free-energy barrier between the states~\cite{peters-book}.   For non-equilibrium systems, dynamics is no longer governed by free-energy gradients, and transitions between metastable states are harder to characterise, although classical nucleation theory can be generalised in some cases, for example active field theories~\cite{cates2023classical}.
The richness of the non-equilibrium barrier-crossing problem is apparent already when considering a single active particle escaping from a potential minimum~\cite{woillez2019activated}.}

{Large deviation theory is the natural mathematical framework for analysing such rare events.   In particular, the theory of Freidlin and Wentzell~\cite{freidlin2012random} describes rare events in dynamical problems involving low-noise (or low-temperature) limits.  A central role is played by the quasipotential, which is the non-equilibrium analog of the energy that appears in the Boltzmann-Gibbs distribution at equilibrium.
The corresponding quasipotential landscape generically features basins of attraction which are connected by transition paths; the rates of transitions between metastable states are determined by differences in the quasipotential between its initial minimum and the transition state {(akin to a saddle point on the landscape)}.  We show in the following that computation of the quasipotential allows the identification and characterisation of Escher cycles.}

The quasipotential of a state can be calculated by identifying the least-unlikely fluctuation path into that state (see below for further details).  In contrast to equilibrium energy landscapes or free-energy surfaces, this is an intrinsically dynamical calculation.  {Such problems are increasingly relevant in non-equilibrium statistical physics, including for chemical reaction networks~\cite{dykman1994large,lazarescu2019large,schuttler2024effects} 
{and active particles \cite{heller2024evaluation,crisanti2023most,yasuda2022most,goswami2023effects}.}}
 It leads to
a striking feature of quasipotentials: they can be non-differentiable, at points where several fluctuation paths are equally likely. A classical example for a particle moving in two-dimensions was given by Jauslin~\cite{jauslin1987nondifferentiable}, and related singularities were noted in mathematical studies of escape problems~\cite{day1987recent}.  Foundational work by Graham and Tél~\cite{graham1985weak,graham1986nonequilibrium}, and by Maier and Stein~\cite{maier1992transition,maier1993escape,maier1993effect} explored such singularities in systems that involve metastability.
{In the present context, these singularities mean that the quasipotential may be non-analytic in the vicinity of the transition state, in contrast to the equilibrium case where transition states are typically saddle points on the {energy (or free-energy) landscape (or surface)}.  We note in passing that the quasipotential is a large deviation rate function, and singularities in such functions are of independent interest in a variety of contexts~\cite{bertini2010lagrangian,baek2015singularities,nemoto2017finite,baek2018dynamical,aminov2014singularities}, including analogies with phase transitions.}

{Given all the above, we emphasize again that the quasipotential is an essential tool for characterisation of rare transitions between metastable states in non-equilibrium systems; it has a rich structure but is challenging to compute in general.  Despite the generic importance of transitions between metastable states, explicit analytic calculations of the quasipotential remain scarce for finite-dimensional systems (notwithstanding the few examples given above).
With this in mind, we present here analytical and numerical calculations a simple non-equilibrium model system that illustrates all these points, including an Escher cycle and a quasipotential that features several types of singularity.}

{We outline our results next, before proceeding to a detailed analysis.}

\subsection{Outline}

We study a system that was motivated by the nonreciprocal Ising model of Avni \textit{et al.}~\cite{avni2023non,avni2024dynamical}.  Specifically, we consider two non-reciprocally coupled variables $x(t),y(t)$ which correspond to two order parameters in a mean-field version of that model.   [In addition to their interpretation as order parameters, it is also sometimes convenient to think of $x(t),y(t)$ as coordinates of a single particle moving in two dimensions.]   The variables $x,y$ follow a dynamics with weak noise, and the system is driven away from equilibrium by non-reciprocal forces.  The system has four metastable states corresponding to $x,y\approx \pm 1$; the system exhibits noise-driven transitions between them in a preferred cyclic order, forming an Escher cycle.  
This situation is similar to the cyclic nucleation phase of the non-reciprocal Ising model~\cite{avni2023non,avni2024dynamical} in which two species of spins undergo a repeating sequence of nucleation events between four differently magnetized states.  Each event is initiated by one of the species, whose ``selfish energy'' is reduced in the transition, so that each step of the Escher cycle is indeed downhill, when viewed from the perspective of the initiating species.

Our analysis proceeds via computation of the quasipotential which we perform in several stages, that also provide interesting intermediate results.  Specifically, the Freidlin-Wentzell theory~\cite{freidlin2012random} is based on a local calculation of the quasipotential, relative to each metastable state.   These relative quasipotentials (RQPs) are then  combined to yield a global quasipotential, which determines the steady state probability distribution.
Both the RQP and the global quasipotential reveal notable (singular) features.  Singularities of the RQP arise when there are degenerate optimal fluctuation paths that lead to the same final point.  In our example, these paths either go directly to the point of interest, or they take a detour through a saddle point.  These fluctuation paths are degenerate on singular lines in the RQP, which separate points where each mechanism is optimal.  While such phenomena have been observed numerically and discussed qualitatively in the past \cite{graham1986nonequilibrium, graham1985weak,maier1992transition, maier1993escape, maier1993effect, bouchet2016perturbative}, this model provides a rare example where analytic calculations can be performed.

The model also supports singularities in the global quasipotential, by a different mechanism.  The global quasipotential is obtained as a pointwise minimum of all RQPs (shifted by appropriate offsets, see Sec.~\ref{sec:glob-defns}).  Physically, this minimisation corresponds to identifying the metastable state in which the optimal fluctuation path originates.  Analogous to the RQP, this leads to singular lines in the global quasipotential, which separate the regions in which each starting point is optimal.
These issues were discussed  also in~\cite{graham1985weak,graham1986nonequilibrium,maier1992transition,maier1993escape,maier1993effect} which used a Hamiltonian formalism to describe the fluctuation paths.  Our approach here uses a Lagrangian picture based directly on the Freidlin--Wentzell action, which aids physical interpretation of the singularities, in terms of local minimisers of the action, and their degeneracies.  The associated variational structure is also exploited by some numerical methods~\cite{grafke2019numerical,kikuchi2020ritz}.

Our results constitute useful progress towards general-purpose tools for analysing rare events in non-equilibrium systems.
The low-dimensional serves as a test bed for methods, including reaction coordinate identification, quasipotential calculation, and optimal path analysis, that are essential for tackling more complex, spatially extended systems. Indeed, the same principles are expected to apply in active matter or ecological models where continuous fields interact via nonreciprocal couplings, but we expect the calculations to be more challenging in that case.

The remainder of the paper is organized as follows. In Section \ref{sec:model}, we define the model and outline its physical features. Section \ref{sec:ldp_qp} introduces the large deviation framework and the concept of the quasipotential, which governs rare-event behaviour in the small-noise limit. In Section \ref{sec:1dqp}, we carry out a one-dimensional reduction of the model and construct the corresponding $1d$ quasipotential in a multistep process, which sets the stage for the full two-dimensional analysis. Section~\ref{sec:perturbative} introduces the methodology for computing the RQP via a perturbative expansion of the Hamilton-Jacobi equation. Section~\ref{sec:main} calculates the full two-dimensional quasipotential in a numerical and perturbative way; we identify competing branches of the quasipotential, compute their action locally with respect to a basin of attraction, and then construct the global quasipotential by stitching these RQPs together. In Section~\ref{sec:asymmetric}, we generalise the model by introducing an extra non-reciprocal coupling parameter, and examine its effect on the quasipotential. 
Finally, we conclude in Sec.~\ref{sec:conclusion}.

\section{Model}\label{sec:model}

We introduce a minimal stochastic model composed of two coupled order parameters $x(t), y(t)$. We write $ \ve{z}(t) = (x(t), y(t)) $ which evolves in time according to
\begin{align}\label{eq:model}
\begin{split}
  \dot{\ve{z}}(t) &= \ve{R}(\ve{z}) + \sqrt{2 \epsilon}\, \bm{\eta}(t) ,
  \\
  \ve{R}(\ve{z}) &= -\nabla U(\ve{z}) + \gamma \bm{\Gamma}(\ve{z}) .
\end{split}
\end{align}
Here, the noise term $\bm{\eta}(t) = (\eta_x(t), \eta_y(t))$ represents Gaussian white noise with zero mean $\langle \eta_i(t)\rangle = 0$ and correlations given by $\langle \eta_i(t)\eta_j(t')\rangle = \delta_{ij}\delta(t-t')$ for $i,j \in \{x,y\}$.
The noise strength $\epsilon$ represents an effective temperature, assumed to be small throughout.
The drift $ \ve{R}(\ve{z}) $ consists of two contributions. The first term, $ -\nabla U(\ve{z}) $, corresponds to deterministic relaxation in a symmetric double-well potential,
\begin{align}\label{eq:U}
  U(\ve{z}) = \frac14 (1-x^2)^2 + \frac14 (1-y^2)^2 ,
\end{align}
which leads to bistability in each variable, analogous to magnetization in an Ising model.
The coupling $ \bm{\Gamma}(\ve{z})$ introduces a nonreciprocal interaction between $ x $ and $ y $. 
We choose
\begin{align}
  \ve{\Gamma}(\ve{z}) = 
  \begin{pmatrix}
    y \\
    -x
  \end{pmatrix} ,
\end{align}
so that the interaction has equal magnitude $| \gamma| $ in both directions but opposite sign. 
In Sec.~\ref{sec:asymmetric}, we generalise this setup to allow for two independent coupling strengths $ \gamma_1 $ and $ \gamma_2 $. {The coupling considered here can be interpreted as $ x $ being ``attracted'' to $ y $, while $ y $ is ``repelled'' by $ x $. The antisymmetry of this coupling in $(x,y)$ space means that it cannot be derived from a scalar potential} and explicitly breaks detailed balance. As a result, the system is driven out of equilibrium whenever $ \gamma \neq 0 $.

\begin{figure}[t]
  \includegraphics{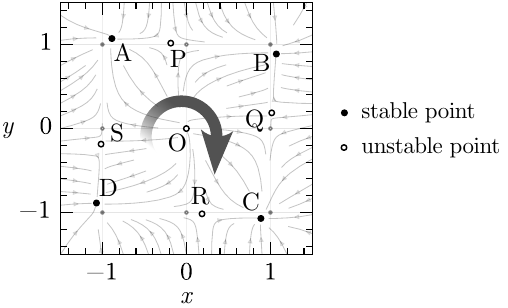}
  \caption{The model in Eq.~\eqref{eq:model} and its fixed points for $\gamma / \gamma_{\rm c} = 0.5$. The thin grey lines show the flow field of the deterministic drift $\ve{R}$. The filled markers show the four stable fixed points $\A,\B,\C,\D$, and the hollow markers show the five unstable fixed points $\mathcal{U} = \{ \sO, \sP, \sQ, \sR, \sS\}$. The smaller grey markers show the equilibrium fixed points for $\gamma = 0$ as a visual guide. The rotating arrow indicates the preferred direction of rotation for positive interaction $\gamma > 0$.}
  \label{fig:model}
\end{figure}

At equilibrium $\gamma = 0$, the dynamics involve nine fixed points $\ve{z}_X$ satisfying $\ve{R}(\ve{z}_X) = 0$: four stable points at the corners of the unit square $(\pm1, \pm1)$ corresponding to the minima of the potential $U$, four unstable points at the midpoints $(\pm1, 0)$ and $(0, \pm1)$, corresponding to saddles of $U$, and one unstable point at the origin $(0, 0)$, corresponding to the maximum in $U$. These are shown as grey markers in Fig.~\ref{fig:model}, with filled circles indicating stable points and hollow circles unstable ones. In the presence of noise, the system typically resides near one of the four stable points, and undergoes rare transitions between basins of attraction, with rates governed by classical Arrhenius theory. The barrier height is $\Delta U = 1/4$, and the transition rate scales as $\exp(-\Delta U / \epsilon)$ as $\epsilon \to 0$, see also Sec.~\ref{sec:ldp_qp}.

For small coupling $0 < |\gamma| < \gamma_{\rm c} = 1/\sqrt{8}$, the system retains the equilibrium topology of fixed points, with four stable points and five unstable ones, but the noise-induced transitions between basins become directionally biased. Specifically, the coupling generates a net circulation in phase space: for $\gamma > 0$, the system exhibits clockwise transitions through the sequence of fixed points, while for $\gamma < 0$ the motion is counter-clockwise. In what follows, we focus on the case $\gamma > 0$ without loss of generality, as illustrated in Fig.~\ref{fig:model}. In this regime, detailed balance is violated, and transition rates can no longer be predicted from $U$ alone. A generalised notion of barrier height is required, which we discuss later using large deviation theory.

When the nonreciprocal coupling surpasses the critical interaction $|\gamma|>\gammac$, a bifurcation occurs in which 
the eight peripheral fixed points (all fixed points but the origin) coalesce to form a stable limit-cycle attractor. The resulting dynamics exhibit sustained periodic oscillations.
This supercritical phase corresponds to the “swap phase” of the nonreciprocal Ising model \cite{avni2023non, avni2024dynamical}, characterised by deterministic cycling of the order parameters.

Thus, the model \eqref{eq:model} captures the two dynamical phases that replace static order in the nonreciprocal Ising model~\cite{avni2023non}: the swap phase, characterised by deterministic oscillations, and the cyclic nucleation phase, marked by noise-driven transitions in a preferred sequence. This makes it a minimal and analytically tractable proxy for isolating the essential mechanisms underlying these behaviours. In particular, we will show that it reproduces the core feature of the Escher cycle observed in the cyclic nucleation phase.

In the remainder of this paper, we focus exclusively on this subcritical regime $0 < \gamma < \gamma_{\rm c}$, where the Escher cycle emerges.
Here, the eight peripheral fixed points are smoothly perturbed from their equilibrium positions. We label the stable fixed points in clockwise order as $X = \A, \B, \C, \D$ and the unstable ones as $X = \sP, \sQ, \sR, \sS$, as shown in Fig.~\ref{fig:model}, where the perturbed fixed points are indicated in black to distinguish from the equilibrium points in grey.
The fixed point $\A$ is located at
\begin{align}\label{eq:x_A}
  \ve{z}_\A = (x_\A, y_\A) \qquad \text{with} \qquad  x_\A = -\frac12 \sqrt{3 + \sqrt{1-8\gamma^2} - \sqrt{2(1+4\gamma^2) - 2\sqrt{1-8\gamma^2} }}, \quad y_\A = \frac{x_\A^3 - x_\A}{\gamma} ,
\end{align}
and the nearby unstable point $\sP$ is given by
\begin{align}\label{eq:x_P}
  \ve{z}_\sP = (x_\sP, y_\sP) \qquad \text{with} \qquad  x_\sP = 
  -\frac12 \sqrt{3 - \sqrt{1-8\gamma^2} - \sqrt{2(1+4\gamma^2) -2 \sqrt{1-8\gamma^2} }}
  , 
  \quad y_\sP = \frac{x_\sP^3 - x_\sP}{\gamma} .
\end{align}
The remaining fixed points are generated by successive $90^\circ$ rotations of these solutions about the origin. 
We denote the stable points (or attractors) by $X \in \{\A,\B,\C,\D\}$, with $\z_X$ the location of the fixed point itself and $\basin_X$ its basin of attraction, i.e. the set of initial conditions that converge to $X$ under the deterministic flow.

In parts of this work, we analyse the system in the weak-coupling regime $\gamma \to 0$. Up to third order in $\gamma$, the fixed points are given by
\begin{subequations}\label{eq:fp_expansion}
\begin{align}
  &x_{\A} = -1 + \frac12 \gamma + \frac58 \gamma^2 + \bigO(\gamma^3),
  \qquad 
  &&y_{\A} = +1 + \frac12 \gamma - \frac58 \gamma^2 + \bigO(\gamma^3),
  \\
  &x_{\sP} = - \gamma + \bigO(\gamma^3),
  \qquad 
  &&y_{\sP} = +1 + \frac12 \gamma^2 + \bigO(\gamma^3),
  \\
  &x_{\B} = +1 + \frac12 \gamma - \frac58 \gamma^2 + \bigO(\gamma^3),
  \qquad 
  &&y_{\B} = +1 - \frac12 \gamma - \frac58 \gamma^2 + \bigO(\gamma^3).
\end{align}
\end{subequations}
We include the expansion for $\B$ to highlight an important asymmetry: we have $x_{\B} - x_{\A} = \bigO(1)$ but $y_\B-y_\A=\bigO(\gamma^2)$. This separation of scales allows us to develop a reduced $1d$ description of the model, see Sec.~\ref{sec:1dqp}.

\section{Large deviation theory and the quasipotential}
\label{sec:ldp_qp}

A central goal of this paper is to understand how nonreciprocal coupling shapes the steady-state distribution $P_{\rm ss}(\z)$ of the system \eqref{eq:model} in the metastable regime, i.e.\ for subcritical coupling $|\gamma|<\gamma_{\rm c}$ and weak noise $\epsilon$. The key questions are how noise drives transitions between metastable states, and how these transitions determine the steady state. 
{Large-deviation theory provides a framework to answer this question, specifically via the theory of Freidlin and Wentzell~\cite{freidlin2012random}: this Section summarises the most relevant parts of that theory.  
It is convenient to interpret the drift $\bm{R}$ as defining a dynamical system: our discussion uses that this has only a finite set of isolated equilibria (fixed points) and no limit cycles (recall that we assumed $|\gamma| < \gamma_c$).}

A central object is the \emph{quasipotential} $F$, which governs rare-event probabilities and encodes the large-deviation structure of $P_{\rm ss}$ in the small-noise limit
\begin{align}\label{eq:Pss_to_F}
  P_{\text{ss}}(\z) \asymp \exp\!\left(-\frac{F(\z)}{\epsilon}\right),
\end{align}
or, equivalently, 
\begin{align}
  F(\bm{z}) := - \lim_{\epsilon \to 0} \epsilon \log P_{\rm ss}(\bm{z}) .
\end{align}
The form \eqref{eq:Pss_to_F} resembles the Boltzmann distribution.  The quasipotential $F$ also plays the role of an effective non-equilibrium potential, in the sense that it decreases monotonically along deterministic relaxation and thus serves as a Lyapunov function, while $\epsilon$ acts as an effective temperature.  For equilibrium systems, which correspond here to $\bm{R}=-\nabla U$, one has $F(\bm{z})=U(\bm{z})-U_0$ with $U_0=\min_{\bm{z}} U(\bm{z})$.

Inserting \eqref{eq:Pss_to_F} as a WKB ansatz into the stationary Fokker--Planck equation associated to Eq.~\eqref{eq:model} and retaining only the leading order in $\epsilon$ yields the stationary Hamilton--Jacobi equation (HJE) \cite{bouchet2016generalisation}
\begin{align}\label{eq:HJE}
  \nabla F(\bm{z}) \cdot \big( \ve R(\z) + \nabla F(\bm{z}) \big) = 0 .
\end{align}
Solving \eqref{eq:HJE} is nontrivial and is typically addressed via characteristics.  {However, this method does not provide a unique solution in non-equilibrium systems, because} multiple characteristics may reach the same point under identical boundary conditions. 
For instance, in the model \eqref{eq:model} for sufficiently large interaction strength $\gamma$, a characteristic following the “around-the-back’’ route $\A\!\to\!\B\!\to\!\C\!\to\!\D$ can compete with a direct route $\A\!\to\!\D$ to the same endpoint $\D$, even if the latter path is shorter. 

The ambiguities are resolved by constructing $F$ from a minimum-action principle, based 
 on the \emph{Freidlin--Wentzell (FW) action}, which is defined for any absolutely continuous trajectory $\varphi:[t_0, t_1]\to\mathbb R^2$ as
\begin{align}\label{eq:action}
  \mathcal S[\varphi]
  := \frac{1}{4}\int_{t_0}^{t_1}\!\big\|\dot\varphi(t)-\ve R\big(\varphi(t)\big)\big\|^2\,\plaind{t} .
\end{align}
The FW large-deviation principle specifies the 
probability that a stochastic trajectory $\z(t)$ remains close to a prescribed path $\varphi(t)$, as
\begin{align}
  P \Big(\z(t) \approx \varphi(t) \Big)
  \asymp
  \exp\!\left(-\frac{\mathcal S[\varphi]}{\epsilon}\right).
\end{align}
Deterministic trajectories $\dot\varphi=\ve R(\varphi)$ have zero action and are therefore ``typical'', whereas deviations cause positive action and are exponentially suppressed. 
{The following Secs.~\ref{sec:rqp-defns} and~\ref{sec:glob-defns} explain how $F$ can be obtained by finding instantons (fluctuation paths) that minimise this action.}

\subsection{Relative quasipotential}
\label{sec:rqp-defns}

{As an intermediate step for the calculation of $F$}, we consider
the \emph{quasipotential relative to a fixed point $X$} (RQP), which is the minimal action required for a fluctuation path to carry the system from the fixed point $\z_X$ to a target point $\z$:
\begin{equation}\label{eq:local_qp}
  F_X(\z) := \inf_{T>0} \min_{\substack{\varphi(-T)=\z_X\\ \varphi(0)=\z}}
    \mathcal S[\varphi]  .
\end{equation}
Clearly $F_X(\z_X)=0$.  Minimisation with respect to the path $\varphi$ also shows that $F_X$ solves the HJE~\eqref{eq:HJE}.

The minimising paths of \eqref{eq:local_qp} are optimal fluctuation paths (\emph{instantons}).
Since $\z_X$ is a fixed point, the optimal transition time $T$ is infinite, and the instanton approaches $\z_X$ exponentially as $t\to -\infty$.  It is therefore natural to view these paths as trajectories on $(-\infty,0]$.
The optimal fluctuation path from $X$ to $\z$ is then
\begin{equation}\label{eq:instanton_def}
  \varphi_{X \to \z}(t)
  := \argmin_{\substack{\varphi(-\infty)=\z_X,\,\varphi(0)=\z}}
  \mathcal S[\varphi].
\end{equation}

As noted above, equilibrium systems have global quasipotential $F(\z)=U(\z)-U_0$.  However, the RQP is still non-trivial in general.  If $X$ is a stable fixed point and $\bm{z}$ is in the basin of attraction of $X$, then the equilibrium RQP is $F^{(0)}_X(\z) = U(\z) - U(\z_X)$.   (Here and below, the zero superscript indicates equilibrium.)  The equilibrium RQP differs from the (global) quasipotential by a constant offset, $U_0-U(\z_X)$.  If $\bm{z}$ is outside the basin of attraction of $X$ then the relationship between the RQP and $U$ is more complicated.  For example, one sees from Eq.~\eqref{eq:action} that the RQP is non-decreasing along minimal-action paths, but this is not the case for $U$.

An important result for the following is that the equilibrium instanton $\varphi_{X\to \z}^{(0)}$ obeys
\begin{align}\label{eq:fluctuation_dynamics_eq}
\begin{split}
  \frac{\plaind}{\plaind{t}}\varphi_{X\to \z}^{(0)}(t) 
  &= \overline{\ve R}^{(0)}_X\!\left(\varphi_{X\to \z}^{(0)}(t)\right), \quad t \le 0,\\
  \lim_{t\to -\infty} \varphi_{X\to \z}^{(0)}(t) &= \z_X^{(0)},\quad 
  \varphi_{X\to \z}^{(0)}(0) = \z,
\end{split}
\end{align}
where
\begin{equation}\label{eq:fluctuation_drift_eq}
  \overline{\ve R}^{(0)}_X(\z) = -\nabla U(\z) + 2\nabla F^{(0)}_X(\z),
\end{equation}
is called the \emph{fluctuation drift}~\cite{bouchet2016generalisation}.   
Note that this quantity depends itself on the RQP whose determination requires the instanton, so~\eqref{eq:fluctuation_drift_eq} is not an explicit solution to the minimisation problem~\eqref{eq:instanton_def}.  However, it is a property of the solution that will be useful in the following.

{For $\z$  within the basin of attraction of $X$, we recall that $\nabla F^{(0)} = \nabla U$ so also $\overline{\ve R}^{(0)}_X(\z)=\nabla U(\z)$.  Comparing with $\ve R^{(0)}(\z)=-\nabla U(\z)$, one sees that the instanton trajectory is related by time-reversal to the spontaneous relaxation trajectory from $\z$ to $X$: this is the Onsager--Machlup symmetry.  Instantons connecting fixed points to positions $\z$ outside their basins of attraction will be discussed for specific cases below.}

Returning to the general (non-equilibrium) case, note that
more than one local minimiser of ${\cal S}$ may connect $X$ to the same target. 
Let $\varphi^k_{X \to \z}$ denote the $k$-th local minimiser:
 the $k$th branch of $F_X$ is then
\begin{equation}
  F_X^k(\z) := \mathcal S\big[\varphi^k_{X \to \z} \big]
  \label{eq:FXk}
\end{equation}
and the RQP at $\z$ is given by their pointwise minimum over branches
\begin{equation}\label{eq:locqp_from_branches}
F_X(\z) = \min_k F_X^k(\z).
\end{equation}
As a pointwise minimum of continuous functions, $F_X$ itself is continuous, but it may lose differentiability at locations where the minimiser switches from one branch to another.

{A comparison of such branches is relevant for the ``round-the-back'' route from A to D in Fig.~\ref{fig:model}, as already mentioned above.  That is, when calculating $F_{\rm A}(\ve{z}_{\rm D})$ for that model, one may construct a path going directly from A to D (via S) that is a local minimum of ${\cal S}$; there is also a locally-minimising path that goes via $B,C$ (and the intervening saddles).
The solution of Eq.~\eqref{eq:local_qp} is obtained by minimising the action over the branches, as dictated by  \eqref{eq:locqp_from_branches}.
The same principle applies when calculating the RQP $F_X(\ve{z})$ for a generic point $\ve{z}$: the instanton which solves Eq.~\eqref{eq:local_qp} may go directly from $X$, or it may pass through one or more intermediate fixed points.  Moreover, the solution of Eq.~\eqref{eq:local_qp} may switch between different branches as $\ve{z}$ is varied, which leads to singularities in $F_X$.  (We have focussed here on branches that involve intermediate fixed points because these will be relevant below; in general there can also be degenerate minimisers of ${\cal S}$ without any intermediate fixed points~\cite{kikuchi2020ritz,baek2015singularities,maier1996scaling}.)}

\subsection{Global quasipotential}
\label{sec:glob-defns}

{Physically, the RQP $F_X(\ve{z})$ measures how unlikely is a fluctuation from $X$ to $\ve{z}$. One interpretation of the global quasipotential $F(\ve{z})$ is that it measures how unlikely is a fluctuation into $\ve{z}$, starting from the steady state. This may be obtained from the RQPs by considering all possible fixed points as origins for the fluctuation, and their relative probabilities as starting points.  To this end}
we introduce for each {stable fixed point} $X$ an \emph{offset} $W_X$ such that 
\begin{equation}\label{eq:W_X}
  P_{\rm ss}(\z_X) \;\asymp\; e^{-W_X/\epsilon}.
\end{equation}
{[so $F(\z_X)=W_X$, by Eq.~\eqref{eq:Pss_to_F}].
The offsets can be computed~\cite{freidlin2012random} by considering a discrete-state Markov chain whose states are the {stable fixed points}, and with transition rates from $X$ to $Y$ given by $k_{X\to Y}={\rm e}^{-F_X(\z_Y)/\epsilon}$ (an example is given in Sec.~\ref{sec:asymmetric}, below).  This steady state of this Markov chain also obeys Eq.~\eqref{eq:W_X} which allows determination of the $W_X$'s.}
The global QP is then assembled by shifting each RQP by its offset, and taking the minimum over all attractors
\begin{equation}\label{eq:global_from_local_qp}
  F(\z) = \min_X\{ F_X(\z) + W_X \} \; . 
\end{equation}

In equilibrium the basin offsets reduce to simple potential differences. Since the stationary distribution is exactly Gibbs, and we consider the zero-temperature limit, Eq.~\eqref{eq:W_X} gives $W_X^{(0)} \approx U(\z_X) - \min_X U(\z_X)$.
The global construction \eqref{eq:global_from_local_qp} then collapses to the familiar equilibrium form (at zero-temperature): $F^{(0)}(\z) = U(\z) - \min_X U(\z_X)$.

{The pointwise minimisation principle \eqref{eq:global_from_local_qp} for the global quasipotential has some similar implications to the minimisation principle \eqref{eq:local_qp} for the RQP: as $\z$ varies, the solution may switch between different branches, leading to singularities. 
From Eq.~\eqref{eq:global_from_local_qp}, $F$ is again continuous as a function of $\z$ but may have discontinuities in its derivatives, which arise when different choices of $X$ are degenerate.  As $\z$ is varied across the discontinuity, the dominant fluctuation path changes its origin from one fixed point to another.}

{To close this Section, we note that our description of the Freidlin--Wentzell theory involves two distinct variational principles, which are Eqs.~(\ref{eq:local_qp},\ref{eq:global_from_local_qp}): these together determine both the RQP and the global quasipotential.  We emphasize that these two minimisations can both lead to singularities in the RQP and global quasipotential.  Singularities in the RQP come from \eqref{eq:local_qp} and involve paths with the same initial and final points, where the behaviour at intermediate times has a discontinuous change as $\z$ is varied.  On the other hand, singularities in the global quasipotential may also originate from step-changes in initial point $X$ of the optimal path Eq.~\eqref{eq:global_from_local_qp}.}
We further distinguish between second-order singularities, where the function remains continuous in the first derivative ($C^1$) but loses $C^2$ smoothness, and first-order singularities, where even the first derivative is discontinuous across the switching set, leaving only $C^0$ continuity.  {Recalling that both $F$ and $F_X$ solve the HJE,} we note that whenever such non-differentiabilities appear in these functions, the HJE must be interpreted in the weak, or viscosity, sense, through comparison with smooth test functions \cite{feng2006large}.

\section{Quasipotential in one dimension}
\label{sec:1dqp}

\begin{figure}[t]
\includegraphics[width=1.0\linewidth]{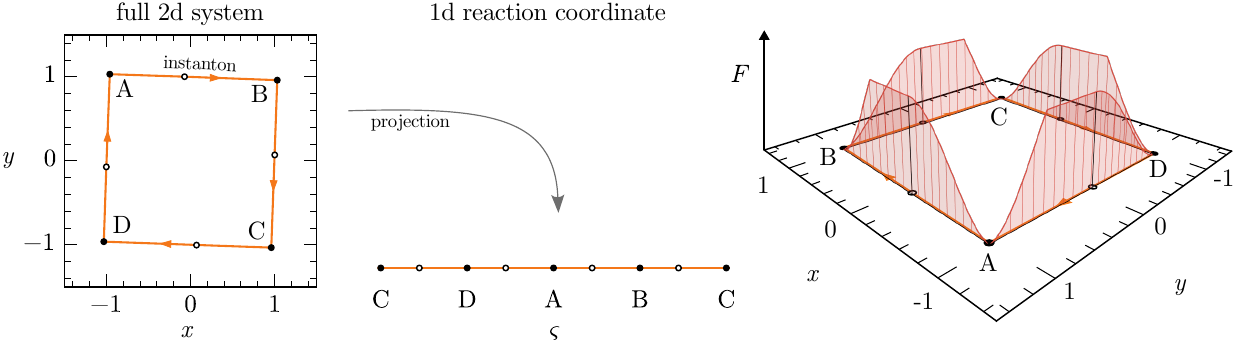}
\caption{
Projection of the $2d$ system onto the $1d$ reaction coordinate $\zo$. 
\textbf{Left}: Instantons (orange) between attractors $\A$, $\B$, $\C$, and $\D$ in clockwise direction for small interaction strength $\gamma$. 
\textbf{Centre}: Reaction coordinate $\zo$ obtained by projecting the instantons from the $2d$ system. Note that the reaction coordinate is periodic. 
\textbf{Right}: Global QP $F(\zo)$ visualised as a height variable above the $1d$ reaction coordinate which lies in the $2d$ configuration space $(x,y)$.
}
\label{fig:1dillustration}
\end{figure}

{In the following we will analyse $F$ and $F_X$ for the model of Fig.~\ref{fig:model}.}
As a warm-up, we first consider projection of the model onto a $1d$ coordinate.  
In this reduced setting the Hamilton--Jacobi equation can be solved explicitly, and instantons can be approximated in a straightforward way, so that both relative and global QPs can be obtained in closed form. 
The example shows how the RQP and global quasipotential can be assembled, as outlined in Sec.~\ref{sec:ldp_qp}.
Although simplified, the $1d$ construction already reveals how nonreciprocity and the nature of the Escher-cycle shape the structure of the QP.

\subsection{Solution of the Hamilton--Jacobi equation in one dimension}

Analysis of the quasipotential is simple in
 $1d$ because the Hamilton--Jacobi equation reduces to an ordinary differential equation with only two possible solution branches.  
Consider {a generic $1d$} system with {periodic} coordinate $\zo$ and drift $r(\zo)$,
\begin{align}\label{eq:projected_dynamics}
  \dot{\zo} = r(\zo) + \sqrt{2\epsilon}\; \xi(t) ,
\end{align}
where $\xi$ is a white noise.
The RQP $F_X(\zo)$ then satisfies
\begin{align}
  F_X'(\zo)^2 + r(\zo)\,F_X'(\zo) = 0 ,
\end{align}
which admits only two pointwise solutions,
\begin{equation}
  F'_X(\zo) =
  \begin{cases}
    0 & \text{(downhill)} ,\\
    -r(\zo) & \text{(uphill)} .
  \end{cases}
  \label{eq:updown}
\end{equation}
The task is therefore to decide, at each position, which of these two branches applies.
Recall, the deciding principle is that $F_X$ solves \eqref{eq:local_qp}.
This leads to a straightforward interpretation:
the instanton from $X$ to $\zo$ generally includes some parts where it moves parallel to $r(\zo)$ (``downhill'') and others where it is antiparallel (``uphill'').  For downhill segments the instanton has $\dot\varphi=r$, the action stays constant along the instanton, and $F_X$ is also constant $F'_X(\zo)=0$.  For uphill segments the solution to  \eqref{eq:local_qp} corresponds to $F'_X(\zo) = -r(\zo)$: the  action accumulates along the instanton.

\subsection{Projection onto $1d$ reaction coordinate}

We now reduce the dynamics among the fixed points of Fig.~\ref{fig:model} to motion along a $1d$ reaction coordinate along which the QP can be calculated explicitly, for small $\gamma$.
This construction is possible because the instantons connecting the fixed points have a simple geometry.

Figure~\ref{fig:1dillustration} shows the projection. The left panel depicts the full dynamics with four attractors $\A,\B,\C,\D$ connected by instantons (orange) in the clockwise direction set by $\gamma > 0$. In the right panel we introduce a $1d$ coordinate $\zo$ that runs along these instantons, mapping each fixed and intermediate point of the $2d$ cycle onto a position on the $\zo$-axis. The full cycle $\A\to\B\to\C\to\D\to\A$ thus becomes a periodic coordinate.  
In this reduced description, advancing in $\zo$ corresponds to following the instanton through successive basins, and wrapping around closes the cycle. 
{The projection into $1d$ greatly simplifies the analysis; the use of periodic boundaries ensures that the ordering of the fixed points is respected.}

Recall that we consider the weak-noise limit $\epsilon \to 0$, and that the interaction strength $\gamma$ is small. To illustrate the construction of the quasipotential, consider the transition $\A \to \B$. In equilibrium, $\gamma=0$, the two degrees of freedom $x$ and $y$ are uncoupled. The transition from $\A$ to $\B$ involves only a change in the $x$-coordinate, since $y_\A = y_\B$. The corresponding instanton is therefore the straight line connecting $x_\A$ and $x_\B$, with the $y$-component fixed throughout.  
When $\gamma$ is nonzero but small, the instanton acquires corrections around the equilibrium path. However, the expansions of the fixed points given in Eqs.~\eqref{eq:fp_expansion} show a scale separation between the two coordinates,  
\begin{equation}
  x_\A - x_\B = \mathcal O(1), 
  \qquad 
  y_\A - y_\B = \mathcal O(\gamma^2) .
\end{equation}
To first order, the $y$-coordinate can therefore be regarded as constant.  
In Fig.~\ref{fig:1dillustration}, this separation of scales manifests itself in the nearly linear form of the instantons, which are simply rotated as a whole around the origin by the first-order corrections to the fixed point coordinates.

Between $\A$ and $\B$, the projection is therefore explicitly defined by setting the reaction coordinate equal to $x$, while freezing the $y$-component at its value in $\A$,  
\begin{equation}
  x \equiv \zo, 
  \qquad 
  y = y_\A = 1 + \bigO(\gamma^2)  .
\end{equation}
The choice of $y=y_\A$ is arbitrary; any constant value along the instanton would lead to the same result to first order.  
As a consequence, the dynamics along the projected coordinate $\zo$ takes the form of Eq.~\eqref{eq:projected_dynamics}, with effective drift
\begin{equation}
  r(\zo) = \zo - \zo^3 + \gamma ,
  \label{eq:rzo}
\end{equation}
for $\zo_{\rm A} < \zo < \zo_{\rm B}$ (with $\zo_{\rm A,B}=x_{\rm A,B}$).
We write $\Delta = \zo_\B - \zo_\A = x_\B-x_\A$ for the distance between adjacent minima.
Fourfold rotational symmetry of Fig.~\ref{fig:model} means that the period of the coordinate $\zo$ is $4\Delta$.

\subsection{Relative quasipotential}

\begin{figure}[t]
\includegraphics[width=0.9\linewidth]{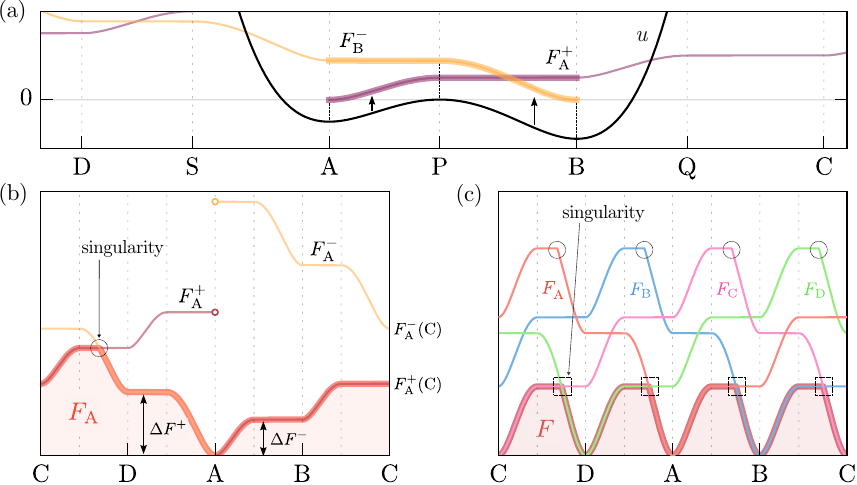}
\caption{
Relative and global QPs in the $1d$ reduction as a function of $\zo$ for $\gamma/\gammac = 0.2$.
\textbf{(a)} Construction of the clockwise (+) and counter-clockwise (-) branches of the RQP $F^\pm_X$. Thick lines indicate initial construction between $\A$ and $\B$, thin line shows the periodic extension. 
\textbf{(b)} Construction of the RQP $F_\A$ (highlighted by a thick light red line with shaded region) as the minimum of both branches. The circle marks a first-order singularity in $F_\A$.
\textbf{(c)} The global QP $F(\zo)$ (shown as a thick dark red line with shading), constructed as the minimum over the four RQPs $F_X(\zo)$ associated with the stable fixed points $X \in \{\A, \B, \C, \D\}$. 
The RQPs $F_X$ for $X = \B, \C, \D$ are obtained from $F_\A$ via translation in space.
The squares mark the first-order singularities in $F$ which are located away from saddles and minima, both of which are marked by the dotted vertical lines. 
}
\label{fig:qp_1d}
\end{figure}

We now construct the RQP $F_{\rm A}(\zo)$ for the full domain of $\zo$ by piecing together {results} from the different regions (which are separated by minima and saddles). 
We begin with the segment between the fixed point $\A$ and the saddle $\sP$. In this region, 
the instanton moving clockwise from $\A$ moves against the drift $r(\zo)$, so we use the uphill solution in \eqref{eq:updown} to obtain 
\begin{align}
  F_\A(\zo) = -\int_{\zo_\A}^\zo r(\zo)\,\mathrm du 
  = u(\zo) - u(\zo_\A), 
  \qquad \zo_\A \leq \zo \leq \zo_\sP ,
    \label{eq:FA-AP}
\end{align}
where we have introduced 
\begin{equation}\label{eq:prepotential}
  u(\zo) = -\int r(\zo)\,\plaind{\zo} 
  = \frac14 (1-\zo^2)^2 - \gamma \zo  + \mathcal O(\gamma^2).
\end{equation}
The resulting $F_\A(\zo)$ and $u(\zo)$ are shown as the black and thick purple lines in Fig.~\ref{fig:qp_1d}(a). 

To obtain the part of $F_\A$ between the saddle $\sS$ and $\A$, we use symmetry arguments. 
So far, our projection was only derived for the segment $\A \to \sP$. 
However, by rotational symmetry in $x,y$ plane [Fig.~\ref{fig:model}], the transition $\A \to \sS$ is equivalent to $\B \to \sP$. Since $\B \to \sP$ is also uphill, repeating the above analysis yields
\begin{align}
  F_\B(\zo) = u(\zo) - u(\zo_\B), 
  \qquad \zo_\sP \leq \zo \leq \zo_\B ,
\end{align}
shown in thick yellow in Fig.~\ref{fig:qp_1d}(a). 
Then  the symmetry implies that 
\begin{equation}
  F_\A(\zo) = F_\B(\zo +\Delta), 
  \qquad \zo_\sS \leq \zo \leq \zo_\A .
  \label{eq:FA-AS}
\end{equation}

Combining (\ref{eq:FA-AP},\ref{eq:FA-AS}) the RQP relative to $\A$ is
\begin{equation}
  F_\A(\zo) = \begin{cases}
    u(\zo) - u(\zo_\A), & \zo_\A \leq \zo \leq \zo_\sP, \\
    u(\zo+\Delta) - u(\zo_\B), & \zo_\sS \leq \zo \leq \zo_\A .
  \end{cases}
\end{equation}
This stitching produces a third-order singularity at $\zo=\zo_\A$, which is due to the ``corner'' at A in the instanton path [Fig.~\ref{fig:1dillustration}(a)]: our focus is on first- and second-order singularities, so we do not analyse higher-order singularities in what follows.
From this expression we can extract the barrier heights to the clockwise and counter-clockwise saddles:
\begin{align}
\begin{split}
  \Delta F^+ &\equiv F_{\A}(\zo_\sP) = u(\zo_\sP) - u(\zo_\A) = \frac14 - \gamma + \mathcal O(\gamma^2), \\
  \Delta F^- &\equiv F_{\A}(\zo_\sS) = u(\zo_\sS) - u(\zo_\A) = \frac14 + \gamma + \mathcal O(\gamma^2).
\end{split}
\end{align}
Here we see the first effect of non-reciprocity: while at equilibrium ($\gamma=0$) the barriers are degenerate, a positive $\gamma$ lowers the clockwise barrier and raises the counter-clockwise one,
\begin{equation}\label{eq:barrier_heights}
  \Delta F^+ < \Delta F^-, \qquad \gamma>0 .
\end{equation}
This asymmetry reflects the intuitive fact that transitions in the direction favoured by the non-reciprocal drive become cheaper.

Next we consider the segment between the saddle $\sP$ and the minimum $\B$.  Assume (see below) that the instanton from $\A$ to any point in this region proceeds via $\sP$.  The action is an integral along the instanton so the quasipotential can be decomposed as 
\begin{equation}
F_\A(\zo) = F_\A(\zo_\sP) + F_\sP(\zo), \qquad  \zo_\sP \leq \zo \leq \zo_\B .
\label{eq:FA-viaP}
\end{equation}  
Moreover, the drift $r$ points away from $\sP$, so one takes the downhill solution in \eqref{eq:updown}, leading to
\begin{equation}
  F_\sP(\zo) = 0 \; ,
  \qquad \zo_\A \leq \zo \leq \zo_\B .
\end{equation}
Then using Eq.~\eqref{eq:FA-viaP} and combining with results above we obtain 
\begin{equation}
  F_\A(\zo) = \begin{cases}
      u(\zo+\Delta) - u(\zo_\B), &  \zo_\sS \leq \zo \leq \zo_\A ,  \\
    u(\zo) - u(\zo_\A), & \zo_\A \leq \zo \leq \zo_\sP, \\
    \Delta F^+, & \zo_\sP \leq \zo \leq \zo_\B ,
  \end{cases}
\end{equation}
which gives the RQP everywhere between $\sS$ and $\B$.

One sees that the RQP develops a flat segment between $\sP$ and $\B$, see Fig.~\ref{fig:qp_1d}(a). This plateau has no counterpart in the underlying potential $u(\zo)$ (or any $1d$ projection of it), but it already appears in the equilibrium RQP $F_A^{(0)}$.
The RQP $F_\A$ generically has a second-order singularity at the transition point $\zo_{\sP}$: its first derivative is continuous [note $u'(\zo_\sP)=0$], but the second derivative is discontinuous [because $u''(\zo_\sP)\neq 0$].  We will encounter more of these features in the following.

To make further progress, we return to the assumption above, that the instanton from A to B proceeds via P.    The instanton from A to a generic point may start by travelling clockwise (via P) or counter-clockwise (via S).  We write $F^+$ for the minimal action along a clockwise path and $F^-$ for the minimal action along an anticlockwise path: these are examples of the ``branches'' that  were anticipated in Eq.~\eqref{eq:FXk}.  Then Eq.~\eqref{eq:locqp_from_branches} becomes
\begin{equation}\label{eq:loc_qp_branches}
  F_\A(\zo) = \min \big\{ F_\A^+(\zo), \, F_\A^-(\zo) \big\} .
\end{equation}
These two branches of the RQP can be computed by following the relevant instanton away from A and using formulae analogous to Eq.~\eqref{eq:FA-viaP}.  For example, suppose that $F^+_\A(\zo_\sQ)$ and $F^-_\A(\zo_\C)$ have already been calculated; then for any point between Q and C, the branches of the RQP obey
\begin{equation}
F_\A^+(\zo) = F^+_\A(\zo_\sQ) + F_\sQ^+(\zo), \qquad F_\A^-(\zo) = F^-_\A(\zo_\C) + F_\C^-(\zo)
\end{equation}
for $\zo_\sQ \leq \zo \leq \zo_\C$.  (Physically, one either travels clockwise to $\sQ$ and then keeps going clockwise from $\sQ$ to the point of interest; or, travel counterclockwise to $\C$ and then further counter-clockwise to the point of interest.)  Calculation of $F_\sQ^+(\zo)$ in this region uses the downhill solution of Eq.~\eqref{eq:updown}, while $F_\C^-(\zo)$ uses the uphill solution.

Figure~\ref{fig:qp_1d}(b) shows both branches as obtained by this construction, together with the resulting $F_\A(\zo)$.  For the generic case ($\gamma\neq0$) there is a single point $\zo^*$ where $F_\A^+=F_\A^->0$.  The RQP $F_\A$ has a first-order  singularity (kink) at this point, which separates two regions where instantons follow either clockwise or counter-clockwise paths.  Note however that in equilibrium ($\gamma=0$), the two branches are degenerate throughout the range $\zo_\sQ < \zo < \zo_\sR$ and there is no kink.

{Physically, a qualitative effect of $\gamma>0$ is that for points between $\zo_\sR$ and $\zo^*$, the instanton from $\A$ travels clockwise, contrary to the equilibrium case where it travelled by the (shorter) counter-clockwise path.  The competition between clockwise and counter-clockwise routes provides a clear example of how nonreciprocity and cyclic topology combine to generate  singularities in the quasipotential landscape.
We will encounter further instances of such singularities when constructing the QP in $2d$.}

\subsection{Global $1d$ quasipotential}
\label{sec:glob-1d}

The global quasipotential $F(\zo)$ in the $1d$ projection is obtained from the RQPs by \eqref{eq:global_from_local_qp}, which we recall amounts to deciding the most likely starting point for a fluctuation to $\zo$.  In the present system the construction is simplified due to symmetry under $90^\circ$ rotations in the $x,y$ plane.  This means that  the offsets $W_X$ in (\ref{eq:W_X},\ref{eq:global_from_local_qp}) are identical and equal to zero for all stable points $X=\A,\B,\C,\D$.  Hence,
\begin{align}\label{eq:F_1dex}
  F(\zo) = \min_{X \in \{\A,\B,\C,\D\}} F_X(\zo) .
\end{align}
By the same symmetry, the RQPs $F_X(\zo)$ for $X=\B,\C,\D$ follow directly from $F_\A(\zo)$ under translation by integer multiples of $\Delta$. 
Physically, the minimisation reflects that the system can arrive at a given $\zo$ via fluctuations from different basins, and with all minima equally weighted in the stationary state, the most probable history dominates.

Figure~\ref{fig:qp_1d}(c) illustrates the resulting global quasipotential, while the right panel of Fig.~\ref{fig:1dillustration} shows the same construction mapped back onto the $2d$ configuration space.
The global quasipotential (red curve with shading) appears as the lower envelope of the four RQPs. 
Because all basins have equal weight, the landscape repeats four times with identical depth, so that the global $F(\zo)$ appears as the same basin pasted around the cycle. 
The structure retains the flat regions beyond saddles already present in the local construction, and acquires a characteristic sawtooth profile: each basin is smooth on its right flank but develops a sharp feature on its left. 

The four kinks [squares in Fig.~\ref{fig:qp_1d}(c)] mark the points where the minimisation in \eqref{eq:F_1dex} switches from one RQP to another.
These are first-order singularities. 
As in the local construction, the equilibrium limit restores symmetry and the corresponding crossovers become second-order, so the appearance of first-order singularities in $F(\zo)$ is a direct signature of nonreciprocity. 
It is important to note, however, that these cusps are distinct from the singularities that arise within the RQPs. 
Indeed, in our example none of the first-order singularities (marked by circles in Fig.~\ref{fig:qp_1d}(c)) of the local construction survive the global minimisation \eqref{eq:F_1dex}; the only singularities in the global QP are those generated by the switching between basins at the global level (squares in Fig.~\ref{fig:qp_1d}(c)).

Thus, even in this reduced $1d$ setting, the global QP displays a rich structure. 
It combines smooth segments with flat regions and singularities that arise at two distinct levels of stitching in the assembly of the local and global QPs. 
In the next section we examine how these features generalise to $2d$. In particular, we will examine how first-order singularities manifest in $2d$ and how the flat regions beyond saddles are reshaped by the nonequilibrium drive.

\section{Perturbative calculation of the relative quasipotential}
\label{sec:perturbative}

We now extend the computation of the RQP from the $1d$ projected manifold to the full $2d$ plane.  The central idea is to expand the RQP around the equilibrium limit, treating the nonreciprocal part of the drift as a small perturbation, following the method of Bouchet et al.~\cite{bouchet2016perturbative}. 
We work here to first order in $\gamma$, which enables some exact results.

We write 
\begin{align}\label{eq:S_expansion}
  F_X(\ve z)
  = F_X^{(0)}(\ve z)
  + \gamma F_X^{(1)}(\ve z)
  + \mathcal O(\gamma^2),
\end{align}
where $F_X^{(0)}$ is the  RQP at equilibrium and the goal is to compute $F_X^{(1)}$.  As illustrated in Sec.~\ref{sec:ldp_qp}, the equilibrium RQP already features singularities: we assume that the instanton path of the equilibrium system is smooth between $X$ and $\ve z$.  This restricts the domain of validity for $\ve z$, but it will be sufficient for the following analysis.

We now recall the method of~\cite{bouchet2016perturbative}, which we present here for the specific case of the first-order perturbative correction.
Inserting the drift $\ve R = -\nabla U + \gamma\,\ve \Gamma$ together with the expansion \eqref{eq:S_expansion} into the HJE \eqref{eq:HJE} and collecting terms of order $\gamma$ yields a first-order equation for $F_X^{(1)}$,
\begin{align}\label{eq:first_order_transport}
  \nabla F_X^{(1)}(\z) \cdot \overline{\ve{R}}^{(0)}_X(\z) = -\nabla F_X^{(0)}(\z) \cdot \ve{\Gamma}(\z)
\end{align}
where $\overline{\ve R}^{(0)}_X(\z)$ is the equilibrium fluctuation drift defined in Eq.~\eqref{eq:fluctuation_drift_eq}.
Now let $\varphi_{X\to \z}^{(0)}(t)$ be an equilibrium instanton connecting $X$ to $\z$, so it solves \eqref{eq:fluctuation_dynamics_eq}. 
We consider how $F_X^{(1)}$ varies along the instanton: 
\begin{align}
\begin{split}
  \frac{\plaind}{\plaind t}\, F_X^{(1)}\!\big(\varphi_{X\to \z}^{(0)}(t)\big)
  &=  \nabla F_X^{(1)}\!\left(\varphi_{X\to \z}^{(0)}(t)\right) \cdot \frac{\plaind}{\plaind t}\varphi_{X\to \z}^{(0)}(t)
  \\
  &=  \nabla F_X^{(1)}\!\left(\varphi_{X\to \z}^{(0)}(t)\right) \cdot \overline{\ve{R}}_X^{(0)}\big( \varphi_{X\to \z}^{(0)}(t) \big)
  \\
  &= -\,\nabla F_X^{(0)}\!\left(\varphi_{X\to \z}^{(0)}(t)\right)  \cdot  \ve{\Gamma}\!\left(\varphi_{X\to \z}^{(0)}(t)\right),
\end{split}
\end{align}
where the first equality is the chain rule, the second is \eqref{eq:fluctuation_dynamics_eq}, and the third is \eqref{eq:first_order_transport}.
Integrating this expression in time 
gives the first-order correction
\begin{align}\label{eq:perturbation_raw}
  F_X^{(1)}(\ve z)
  = F_X^{(1)}(\ve{z}_X^{(0)}) - \int^0_{-\infty} \nabla F_X^{(0)} \left(\varphi_{X\to \z}^{(0)}(t)\right)
     \cdot \ve \Gamma \left(\varphi_{X\to \z}^{(0)}(t)\right)\,\mathrm dt ,
\end{align}
where $\ve{z}_X^{(0)}$ is the position of the fixed point $X$ in the equilibrium case.
Note however that $F_X(\ve{z}_X)=0$ by definition [where $\ve{z}_X$ is the $\gamma$-dependent fixed point], and that $F_X$ is quadratic near $\ve{z}_X$, and that $\ve z_X=\ve z_X^{(0)}+\mathcal O(\gamma)$, which means that $F_X(\ve{z}_X^{(0)})=O(\gamma^2)$ and hence $F_X^{(1)}(\ve{z}_X^{(0)})=0$.

Therefore, combining \eqref{eq:S_expansion} and \eqref{eq:perturbation_raw}
\begin{align}\label{eq:perturbation}
  F_X(\ve z)
  = F_X^{(0)}(\ve z)
    - \gamma \int_{-\infty}^{0} \nabla F_X^{(0)} \left(\varphi_{X\to \z}^{(0)}(t)\right)
     \cdot \ve \Gamma \left(\varphi_{X\to \z}^{(0)}(t)\right)\,\mathrm dt
    + \mathcal O(\gamma^2)
\end{align}
gives the RQP up to first order.
Importantly, this expression depends only on \emph{known} quantities: an equilibrium instanton $\varphi_{X\to \z}^{(0)}(t)$, the equilibrium RQP $F^{(0)}_X$, and the first-order part of the drift $\ve{\Gamma}$.  Recall, its validity relies on the instanton being smooth between $X$ and $\ve z$.
It also relies on the validity of the perturbative expansion in Eq.~\eqref{eq:S_expansion}.
To see the limitations that this entails, note that Eq.~\eqref{eq:S_expansion} is formally valid for both signs of $\gamma$ \cite{bouchet2016perturbative} but the results of Figs.~\ref{fig:1dillustration},\ref{fig:qp_1d} show profound asymmetries in the response to $\gamma$, depending on its sign.  These effects occur near singular points of $F_X$, where the radius of convergence of the perturbation expansion vanishes.  Nevertheless, the following Section shows how accurate approximations for $F_X$ can be obtained at small finite $\gamma$, by stitching together RQPs relative to different fixed points.

\section{Quasipotential in two dimensions}\label{sec:main}

In this section, we calculate the full $2d$ quasipotential $F(\ve{z})$ associated with the stochastic dynamics in Eq.~\eqref{eq:model} to leading order in $\gamma$, and we compare the results with numerical calculations obtained at finite $\gamma$.
The $1d$ calculation in Sec.~\ref{sec:1dqp} provided an analytically tractable example of how the quasipotential is constructed in a system with multiple metastable states. There, the quasipotential was built in three steps: first by identifying two distinct branches $F^\pm_\A(\zo)$ of the RQP  associated with the action in clockwise and counter-clockwise direction; next by taking their minimum to define the RQP $F_\A(\zo)$; and finally by assembling the global quasipotential $F(\zo)$ as the minimum over RQPs associated with each metastable basin. This three-step procedure is also used in the full $2d$ setting.

\begin{figure}[t]
  \includegraphics{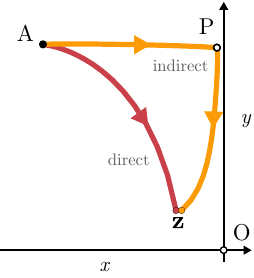}
  \caption{Direct vs. indirect stationary paths in the $2d$ system connecting the stable fixed point $\A$ with targets $\ve{z}$.  These are the instantons that were computed numerically for $\gamma/\gamma_c=0.1$ and $\ve z = (-0.243, 0.2)$ and $(-0.242, 0.2)$. The direct path connects $\A$ and $\ve{z}$ directly, while the indirect path passes through the unstable fixed point $\sP$ on the way.  {For each of the two points, direct and indirect paths both exist as local minima of the action; we only show the global minimum in each case, to illustrate its switch between the branches.}
}
  \label{fig:directvsindirect}
\end{figure}

The key difference in $2d$ is that there is a greater diversity of paths connecting a point $\ve{z}$ to a fixed point $X$.  (In $1d$, only clockwise and counter-clockwise paths were considered, but this is not sufficient for $2d$.)  However, it still suffices to consider two types of path, from which the quasipotential can be computed, to first order in $\gamma$.  
The first type of path (which we call ``direct'') connects the initial fixed point to the target point along a smooth trajectory. The second type (``indirect'') visits one or more other fixed points along the way, resulting in a two-stage instanton.  An example is given in Fig.~\ref{fig:directvsindirect}. These different paths define two separate branches of the RQP [recall Eq.~\eqref{eq:FXk}].

\subsection{Step 1: branches of the relative quasipotential}
\label{sec:branches}

\subsubsection{Direct branch}
\label{sec:direct_branch}

We first compute the direct branch RQP starting from fixed point A.  
(Due to the underlying $\mathbb{Z}_4$ symmetry of the system, the remaining RQPs $ F_X(\ve{z}) $ for $ X = \B, \C, \D $ can then be obtained by $90^\circ$ rotations.)
For the equilibrium system, it turns out that direct paths only exist for points $\ve z$ in the basin of attraction of point A.  We denote this by 
$\basin_\A^{(0)} = \{(x,y) : x < 0, y > 0\}$.  
Recall that for $\ve z \in \basin_\A^{(0)}$, the equilibrium RQP $F_\A^{(0)}$ is equal to the potential $U$ defined in Eq.~\eqref{eq:U}, as discussed in Sec.~\ref{sec:ldp_qp}.
The corresponding equilibrium instanton $\varphi_{\A \to \ve{z}}^{(0)}(t)$ solving  Eq.~\eqref{eq:fluctuation_dynamics_eq} can be calculated exactly because the drift is separable in $x$ and $y$ at $\gamma = 0$: it is given by
\begin{equation}\label{eq:instanton_zI}
  \varphi_{X\to \z}^{(0)}(t) = \begin{pmatrix}
    -\varrho(t,x)
    \\
    +\varrho(t,y)
 \end{pmatrix}
\end{equation}
with
\begin{equation}\label{eq:inst_phi}
  \varrho(t,s) = \frac{|s|}{\sqrt{s^2 + \left(1 - s^2\right) e^{2t}}} .
\end{equation}
This path originates from the correct equilibrium fixed point: $\lim_{t\to -\infty}\varphi_{\A \to \z}^{(0)}(t) = \ve{z}_\A^{(0)} = (-1,1)$ for any $\z \in \basin_A^{(0)}$. 

\begin{figure}[t]
  \includegraphics[width=\linewidth]{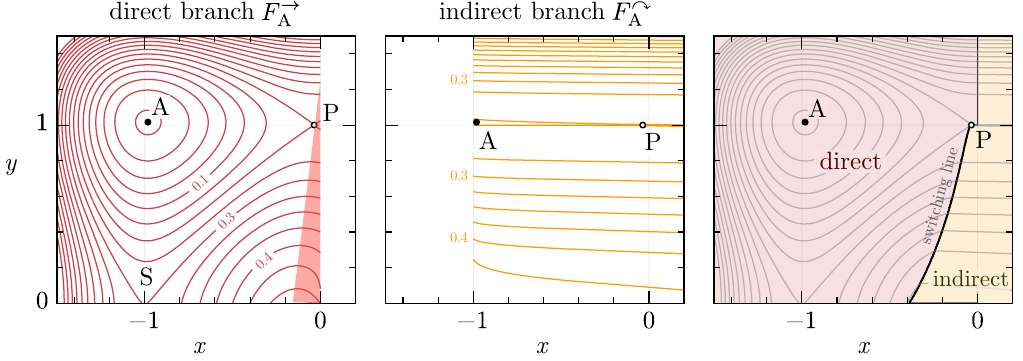}
\caption{
Construction of RQPs based on the two branches of the RQP for $\gamma = 0.1\gammac$.
The contours are equally spaced between $0.05$ and $0.6$, and include the saddle contour $\frac14 - 0.1 \gammac \approx 0.216$.
\textbf{Left}: Direct branch of the RQP. The red shaded region highlights parts of the direct branch that break monotonicity in $x$-direction.
\textbf{Centre}: Indirect branch of the RQP.
\textbf{Right}: The RQP $F_\A$, showing the regions where it derives from each branch. 
The black line marks the boundary where the instanton path switches from direct to indirect. The thick part of this line highlights the first-order part region where indirect and direct paths are distinct, see also Fig.~\ref{fig:path_dissimilarity}.
}
\label{fig:qp_branches}
\end{figure}

Substituting $\nabla F^{(0)}_\A = \nabla U$ and \eqref{eq:instanton_zI} into the integral expression~\eqref{eq:perturbation}, we obtain the first-order correction to the QP
\begin{align}
   F_\A^{(1),\rightarrow}(\ve{z}) = - \int_{-\infty}^0 \nabla U \left(\varphi_{\A \to \z}^{(0)}(t) \right) \cdot \ve{\Gamma}\left( \varphi_{\A \to \z}^{(0)}(t) \right)\,\mathrm{d}t = 
   \frac{(x+y)(1-xy)}{x-y} ,
\end{align}
where the $\rightarrow$ indicates the direct branch.
Using \eqref{eq:perturbation}, we obtain the direct branch of the RQP to first order in $\gamma$ for points $\ve{z} = (x,y) \in \basin_\A^{(0)}$ :
\begin{align}\label{eq:F_A_branch1}
  F_{\A}^\rightarrow(\ve{z}) = \frac{1}{4}(1 - x^2)^2 + \frac{1}{4}(1 - y^2)^2 + \gamma \frac{(x+y)(1-xy)}{x-y} + \bigO(\gamma^2) .
\end{align}
This expression is plotted in the left panel of Fig.~\ref{fig:qp_branches}.
Compared to the equilibrium potential $U$ the contours of the perturbed QP are ``stretched'' in the clockwise direction, resulting in a lower action at saddle $\sP$ (located on the contour $F_X=\frac{1}{4} - \gamma \approx 0.22 $) compared to the equilibrium case with barrier $0.25$, and compared to the saddle $\sS$ (on the contour $F_X=\frac{1}{4} + \gamma \approx 0.28$), as evident from the contour lines. 

To interpret the left panel of  Fig.~\ref{fig:qp_branches}, it is important to remember that the $\bigO(\gamma^2)$ corrections in \eqref{eq:F_A_branch1} have been dropped, and a finite value of $\gamma$ has been substituted.  The accuracy of the resulting estimates of $F_{\A}^\rightarrow$ relies on smallness of higher-order terms.  We show later (Fig.~\ref{fig:qp_local}) that the estimates are accurate across most of $\basin_\A^{(0)}$.  However, there are regions near the edge of the basin where the perturbation expansion breaks down.  This phenomenon is discussed in Appendix~\ref{app:failure}, using calculations for simpler models.  In Fig.~\ref{fig:qp_branches}, the signature of this breakdown appears in the red shaded region, where $F_{\A}^\rightarrow$ is non-monotonic in $x$; this is associated with situations where the quasipotential decreases along an instanton, which is in contradiction with~\eqref{eq:action}.   To avoid artefacts arising from this fact, we explain in Appendix~\ref{app:branch_to_local} that we adjust the perturbative estimate of $F^\rightarrow_\A$ when constructing the RQP.

\subsubsection{Indirect branch}
\label{sec:indirect_branch}

We now compute the indirect branch of the RQP, denoted $ F_{\A}^{\curvearrowright}(\ve{z}) $. Specifically, we consider paths that travel (directly) from A to the unstable fixed point $\sP$ and then go directly to $\ve z$.  Hence,
\begin{align}\label{eq:F_A_indirect}
  F_{\A}^{\curvearrowright}(\ve{z}) = F_{\A}^{\rightarrow}(\ve{z}_\sP) + F_{\sP}^{\rightarrow}(\ve{z}).
\end{align}
From \eqref{eq:F_A_branch1}, the action for the direct path to $\sP$ is $F_{\A}^{\rightarrow}(\ve{z}_\sP)=\frac14 - \gamma + \bigO(\gamma^2)$.  [Continuity of the RQP means that \eqref{eq:F_A_branch1} is valid at $\ve{z}_\sP$, even though this point is not in the interior of $\basin_\A^{(0)}$.]
To compute $ F_{\sP}^{\rightarrow}(\ve{z}) $, we proceed analogously to the direct branch. At $\gamma = 0$, the equilibrium instantons satisfying Eq.~\eqref{eq:fluctuation_dynamics_eq} are 
\begin{equation}\label{eq:instanton_zI_saddle}
  \varphi_{\sP \to \z}^{(0)}(t) = \begin{pmatrix}
    \sign(x) \varrho(-t,x)
    \\
    \varrho(t,y)
 \end{pmatrix},
\end{equation}
valid for $\ve{z} \in \basin_\sP^{(0)} = \{(x,y) :  |x| < 1, y > 0\}$.
Note the $x$-component is a relaxation trajectory (``downhill'', as indicated by the change in sign of $t$) so it does not incur any contribution to the action $S$.  Hence,
\begin{equation}\label{eq:F0_saddle}
  F^{(0)}_\sP(\ve{z}) = \frac1{4} (1 - y^2)^2 ,
\end{equation}
for $\ve{z} \in \basin_\sP^{(0)}$. 
Inserting the instanton \eqref{eq:instanton_zI_saddle} into the integral expression \eqref{eq:perturbation}, we obtain the RQP at first-order 
\begin{align}\label{eq:branch_direct_saddle}
  F_{\sP}^{\rightarrow}(\ve{z}) = \frac{1}{4}(1 - y^2)^2 - \gamma xy(1-y^2) \int_0^1 \frac{u^2 \plaind{u}}{[(1-x^2)+x^2u^2]^{1/2} \, [y^2 + (1-y^2)u^2]^{3/2}} + \bigO(\gamma^2) .
\end{align}
The integral can be expressed in terms of analytic continuations of elliptic functions, but we retain this form here, to emphasize that it is real and positive for all $\ve{z} \in \basin_\sP^{(0)}$.

The centre panel of Fig.~\ref{fig:qp_branches} shows $F_{\A}^{\curvearrowright}$. 
For $\gamma=0$ the contours would be horizontal (parallel to the $x$ axis) but for this positive value of $\gamma$ they tilt slightly clockwise, with the effect becoming more pronounced in the lower-left part in Fig.~\ref{fig:qp_branches}(centre), where both $x$ and $y$ are small.

As a final remark for this section, note that the simple forms of Eqs.~\eqref{eq:instanton_zI_saddle}  and \eqref{eq:F0_saddle} rely crucially on a structural property of our model: for $\gamma=0$, the $x$ component of the gradient  $\nabla U$ only depends on $x$ (and similarly for $y$). This separability allows the decoupling of the equations and leads to closed-form expressions. Such simplifications are not generic and should not be expected in arbitrary $2d$ systems, even in equilibrium. Therefore, the tractability here stems not merely from perturbing around an equilibrium setup; it also relies on specific features of the model.

\subsection{Step 2: Relative quasipotential}
\label{sec:local_qp}

\begin{figure}[t]
\includegraphics[width=\linewidth]{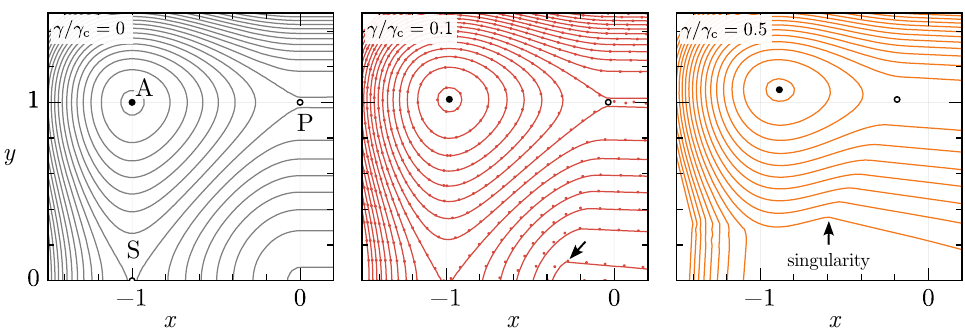}
\caption{
Numerical evaluation of the RQP $F_\A$ for three values of interaction strength $\gamma/\gamma_{\rm c} = 0, 0.1, 0.5$. The left panel with $\gamma/\gamma_{\rm c} = 0$ corresponds to the equilibrium baseline case, the centre panel with $\gamma/\gamma_{\rm c} = 0.1$ showcases the limit of small interaction in which we compare to analytical results (red circles), and the right panel with $\gamma/\gamma_{\rm c} = 0.5$ shows an example of a large interaction, where in the bottom region a plateau of the RQP is visible as a large white region. 
Arrows exemplify points on the contours shown where the RQP developed non-differentiable points as a result of minimisation in Eq.~\eqref{eq:Fm_from_stationarypaths}.
The contours are equally spaced between $0.05$ and $0.6$ and are passing through the value $\frac14 - 0.1 \gammac \approx 0.216$.
}
\label{fig:qp_local}
\end{figure}

\begin{figure}[t]
\includegraphics[width=\linewidth]{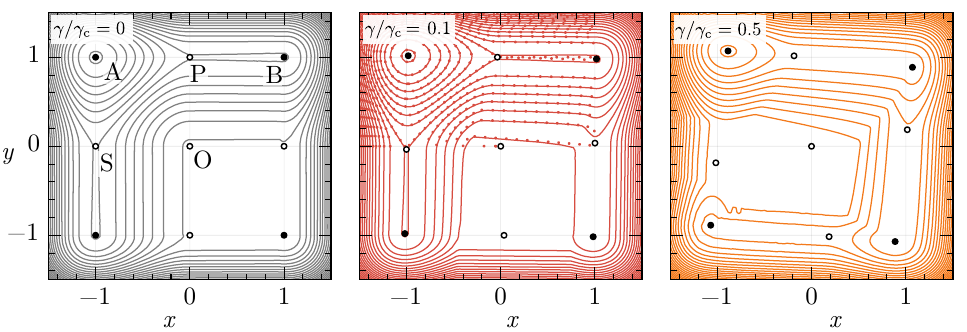}
\caption{
The same numerical evaluation of the RQP $F_\A$ as in Fig.~\ref{fig:qp_local} but shown on a larger domain $[-1.5,1.5]^2$. 
}
\label{fig:qp_local_full}
\end{figure}

\subsubsection{Construction of RQP}

We now construct the relative quasipotential $ F_\A(\ve{z}) $ for the stable fixed point $ \A $. As discussed in Sec.~\ref{sec:ldp_qp}, the RQP is given by the minimum action required to reach $ \ve{z} $ from $\A$.
For the combined domain $\basin_\A^{(0)} \cup \basin_\sP^{(0)} = \{(x,y)  :  y>0, x<1\}$, this means
\begin{align}\label{eq:Fm_from_stationarypaths}
  F_{\A}(\ve{z}) = 
    \min\big\{ F_{\A}^\rightarrow(\ve{z}), F_{\A}^\curvearrowright(\ve{z}) \big\} .
\end{align}
Note however that $F_{\A}^\rightarrow(\ve{z})$ is only defined on $\basin_\A^{(0)}$, similarly $F_{\A}^\curvearrowright(\ve{z})$ is defined only on $\basin_\sP^{(0)}$.  Hence, the minimisation is non-trivial only in $\basin_\A^{(0)} \cap \basin^{(0)}_\sP$; if only one of the two objects is defined then this is the value that is taken.  Also, recall that the method of Appendix~\ref{app:branch_to_local} is used to avoid artefacts associated with the breakdown of perturbation theory for $F_{\A}^\rightarrow$.

The resulting RQP $F_\A$ and the regions where each branch dominates the minimisation \eqref{eq:Fm_from_stationarypaths} are shown in the right panel of Fig.~\ref{fig:qp_branches}. 
(These are the analytical results obtained by the perturbative expansions of the two branches.)
The red and orange regions indicate where the direct and indirect paths are optimal.
The boundary between these regions marks a switch in the dominant fluctuation path which we discuss below. 
{Comparing Figs.~\ref{fig:qp_branches} and~\ref{fig:qp_local} one sees that the indirect branch is the relevant one throughout the (red shaded) region where the adjustment of Appendix~\ref{app:branch_to_local} is applied.  Since the adjustment only affects the direct branch, this means that the results are not sensitive to fine details of the adjustment procedure.}

To complement these analytical results, we also performed numerical calculations using the method of~\cite{kikuchi2020ritz}, see Appendix~\ref{app:numerics} for details.
Fig.~\ref{fig:qp_local} shows $F_\A$ for increasing values of the nonreciprocal interaction $\gamma$. The plots illustrate how the RQP is progressively deformed.
In particular, the barrier is lowered along the direction of positive $x$, consistent with the clockwise rotational bias induced by $\gamma > 0$. 
This deformation favours fluctuation pathways aligned with the non-equilibrium circulation.

The centre panel of Fig.~\ref{fig:qp_local} compares the perturbative analytical result for $F_\A$ (markers) with the numerical results (lines) for small nonreciprocal interaction strength $\gamma = 0.1 \gamma_{\rm c}$. In this regime, the agreement between analytical and numerical results is excellent across most of the domain. Both the shape of the contours and the location of the switching boundary between direct and indirect branches are captured with high accuracy. Notably, the singularity along this boundary appears in both methods, confirming that the perturbative expansion captures not only smooth features but also nontrivial structural elements of the QP landscape. Furthermore, the analytical result also resolves the infinitesimally narrow escape channel to the right of the saddle $\sP$, which the numerical method cannot resolve.

Discrepancies become more pronounced near the centre of the domain, particularly around $x = 0$ and $y = 0$. While numerics and analytics still agree qualitatively, they are slightly offset.  
{As explained above, the behaviour near $x=0$ is associated with the breakdown of perturbation theory, so this is not surprising.  Even so, the agreement between perturbative and numerical results is still semi-quantitative.}

As in the $1d$ case, the RQP $F_\A$ can be extended further by stitching together additional branches to yield a global covering of the full $(x,y)$ space. The outcome of this construction is shown in Fig.~\ref{fig:qp_local_full}.  
In both Figs.~\ref{fig:qp_local} and \ref{fig:qp_local_full}, the flat segments identified in the $1d$ case (Sec.~\ref{sec:1dqp}) persist in $2d$, now appearing as flat regions and extended plateaus. In equilibrium (left panels of Figs.~\ref{fig:qp_local},\ref{fig:qp_local_full}), flat patches emerge beyond the saddles $\sS$ and $\sP$, while a plateau forms downstream of the maximum $\sO$. 
As in the $1d$ case, these features can be understood through the direct/indirect path picture: once the system has paid the cost to reach a saddle, motion along its downhill directions does not accumulate further action to leading order, so the RQP remains constant. The key difference in $2d$ is that saddles have both stable and unstable directions. As a result, the RQP is flat only along the unstable direction, while it retains curvature in the stable one.
Out of equilibrium, the same structures persist but are sheared by non-reciprocity (right panels of Figs.~\ref{fig:qp_local},\ref{fig:qp_local_full}) and acquire directional anisotropy that couples the two coordinates.

\begin{figure}[t]
\includegraphics[width=0.4\linewidth]{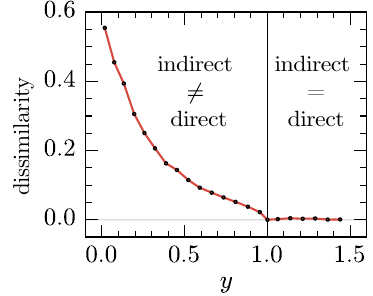}
\caption{
Dissimilarity of the direct and indirect paths along the singular line shown in the right panel of Fig.~\ref{fig:qp_branches} (black line) as a function of the coordinate $y$, for $\gamma = 0.1 \gammac$. This is quantified by the path dissimilarity $\int \| \phi^{\rm direct}_\A(t) - \phi^{\rm indirect}_\A(t) \| \,\mathrm{d}t$, where $\|\cdot\|$ denotes the Euclidean distance. The result shows that the cusp becomes weaker near the saddle at $y \approx 1$, where the two trajectories coalesce, while away from the saddle the growing deviation between the two paths manifests as a sharper cusp in the contour structure evident in the right panel of Fig.~\ref{fig:qp_branches}. 
}
\label{fig:path_dissimilarity}
\end{figure}

\subsubsection{Singularities of $F_\A$}

Recall from Sec.~\ref{sec:1dqp} that the RQP exhibits singularities when projected along the one-dimensional periodic coordinate.  The singularities are second-order for the equilibrium system ($\gamma=0$) and otherwise first-order {(the distinction here is whether the gradient of the quasipotential is continuous at the transition, recall Sec.~\ref{sec:glob-defns}).}
In the left panels of Figs.~\ref{fig:qp_local},\ref{fig:qp_local_full} (equilibrium cases), the second-order singularities appear in the contour lines as junctions between curved and straight line segments.  The first-order singularities are visible as the thick black line in the right panel of Fig.~\ref{fig:qp_branches} and, for illustration, marked by arrows in Fig.~\ref{fig:qp_local} (though additional unmarked occurrences are also present).   Along these lines the QP remains continuous but loses differentiability, which reflects a change in the dominant path: on one side the transition proceeds directly, while on the other it chooses the indirect path via the saddle.

For the first-order case, the singular features in the contour lines become less pronounced as one approaches the saddle point P.  This reflects that the direct and indirect paths become increasingly similar on approaching this point.   This effect is quantified Figure~\ref{fig:path_dissimilarity}.  (It is natural because the indirect path is defined to take a detour via P, but the size of this detour is very short if the final point is itself close to P.)  Moreover, for $y>1$ the direct equilibrium instanton already passes via the saddle, so the two branches merge with each other.
Mathematically, the switching line changes character at this point: it represents a nonequilibrium, first-order singularity before the saddle, but it becomes a second-order singularity after the saddle.

\subsection{Step 3: Global quasipotential}
\label{sec:global_qp_construction}

We finally assemble the global quasipotential $F(\ve{z})$, which encodes the full landscape of rare-event probabilities across the entire state space. To achieve this, we combine the RQPs $F_X(\ve{z})$ associated with each stable fixed point $X \in \{\A, \B, \C, \D\}$ using the general assembly rule defined in Eq.~\eqref{eq:global_from_local_qp}.
As already explained in Sec.~\ref{sec:glob-1d} the basin offsets are equal to each other by symmetry, hence they must all be zero, and we have
\begin{equation}\label{eq:F_min_symmetric}
  F(\ve{z}) = \min_{X \in \{\A,\B,\C,\D\}} F_X(\ve{z}) . 
\end{equation}

\begin{figure}[t]
\includegraphics[width=\linewidth]{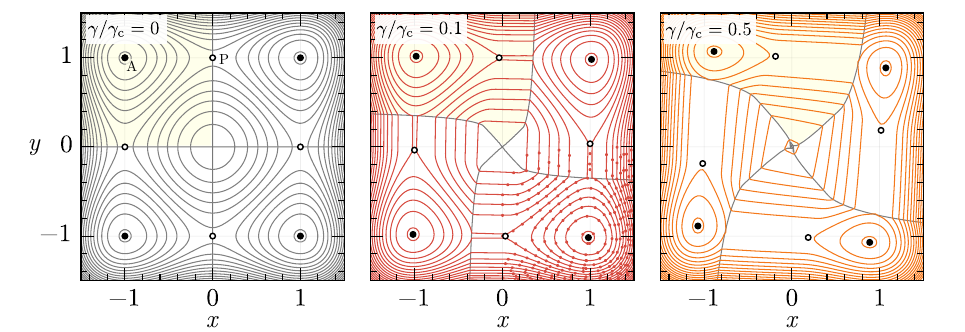}
\caption{
Evaluation of the global quasipotential $F$ computed numerically and analytically for three interaction strengths:
(Left) The equilibrium baseline, $\gamma = 0$.
(Centre) Small non-reciprocity. $\gamma/\gamma_{\rm c}=0.1$. 
(Right) Moderate non-reciprocity, $\gamma/\gamma_{\rm c}=0.5$.
Gray lines mark the switching lines $\Sigma_{\mathrm{bs}}$ where the minimising RQP in Eq.~\eqref{eq:F_min_symmetric} changes. The yellow shading highlights the region in which $F_\A$ is the local minimiser of Eq.~\eqref{eq:F_min_symmetric}, i.e. all $\ve{z}$ for which $F(\ve{z}) = F_\A(\ve{z})$. 
Solid and open markers indicate stable fixed points and unstable fixed points, respectively.
Red dots in centre panel shows analytical perturbative results up to first order in $\gamma$, demonstrating good agreement with numerics. }
\label{fig:qp_global}
\end{figure}

Figure~\ref{fig:qp_global} shows the numerically computed global quasipotential $F(\ve{z})$ for the same interaction strengths as in Fig.~\ref{fig:qp_local}.  (This is obtained from the numerical calculation of $F_\A$, as explained in Appendix~\ref{app:numerics}.)
As in Fig.~\ref{fig:qp_local}, increasing the interaction strength visibly deforms the contours by stretching the boundaries between RQP marked in grey clockwise around the origin.  The central panel of Fig.~\ref{fig:qp_global} also shows analytical results from perturbation theory (as points); these match well with the numerics, similar to Figs.~\ref{fig:qp_local},\ref{fig:qp_local_full}.

The gray lines in Fig.~\ref{fig:qp_global} indicate switching boundaries where the minimiser in Eq.~\eqref{eq:F_min_symmetric} changes between different RQPs. In the yellow region, for example, the global quasipotential coincides (up to a constant) with the RQP associated with the stable point $\A$, and analogous regions exist for the other stable fixed points. At these boundaries the global quasipotential develops non-differentiable features, reflecting the competition between basins of origin: for a point in the yellow region, among all instantons arriving from different fixed points, the dominant contribution is the fluctuation starting at $\A$.

As a consequence, the global quasipotential exhibits singularities on two structural levels. The first level arises within each RQP, when the optimal fluctuation path from a given fixed point switches between direct and indirect escape routes, producing non-differentiabilities in $F_X(\ve{z})$. The second level emerges in the assembly of the global quasipotential through the pointwise minimisation over local branches: wherever the minimising $F_X$ in \eqref{eq:F_min_symmetric} changes, additional switching singularities appear that separate the basins of attraction. 

These levels reflect two features of the model, which 
the existence of several fixed points of the dynamics (multiple metastable states) and the presence of non-equilibrium driving forces.  The first feature necessitates patching together RQPs to yield the global quasipotential; the second means that this quasipotential does not coincide with the (smooth) underlying potential energy surface, so that singularities may appear.  The same two features are responsible for the Escher cycle, which also requires that two (or more) of the metastable states have similar statistical weights in the invariant measure.\footnote{If one state is overwhelmingly dominant in the invariant measure then rare excursions to other states may still take place by mechanisms that violate time-reversal symmetry, but we do not call these Escher cycles.}
One should also distinguish two types of non-equilibrium forcing: it is possible to stir the system locally within its basins without affecting the quasipotential~\cite{spilio2016,kaiser2017acceleration}; the Escher cycle relies on a forcing that affects transitions between basins, leading to a global circulation between then.  We expect that generic non-equilibrium systems will feature both global and local circulation, in which case one may expect Escher cycles to occur, given the conditions above (multiple metastable states with comparable weights, and nonequilbrium forcing).

\section{Reduced symmetry case} 
\label{sec:asymmetric}
\newcommand{\ogg}{\bigO(\gamma_1^2,\gamma_2^2,\gamma_1\gamma_2)}

Until now, our analysis has focused on a system with a single non-reciprocal coupling constant $\gamma$, and the noise terms were equal for both coordinates $x,y$, giving rise to fourfold rotational ($\mathbb{Z}_4$) symmetry. This leads to identical offsets $W_X$ for all stable fixed points $X \in \{ \A, \B, \C, \D \}$. It is possible to break this symmetry by introducing asymmetries in either the interactions~\cite{avni2023non,avni2024dynamical} or the noise.
We implement this by writing
\begin{subequations}\label{eq:model_asymmetric}
\begin{align}
\dot{x} &= x - x^3 + \gamma_1 y + \sqrt{2\epsilon D_1} \, \eta_1(t) ,
\\
\dot{y} &= y - y^3 - \gamma_2 x + \sqrt{2\epsilon D_2} \, \eta_2(t) ,
\end{align}
\end{subequations}
where the parameters $\gamma_1 , \gamma_2$ are the couplings (which are non-reciprocal in general); also  $D_1 , D_2$ multiply the noise acting on each coordinate. Either source of asymmetry ($\gamma_1\neq\gamma_2$ or $D_1\neq D_2$) reduces the rotational symmetry of the system from fourfold to twofold, breaking the degeneracy between metastable states and leading to unequal basin offsets.  
Note that for $\gamma_1 D_2 + \gamma_2 D_1=0$  the couplings are in fact reciprocal, and the system has detailed balance, with a Boltzmann steady state and quasipotential $F(x,y) = (x^2-1)^2/(4D_1) - (y^2-1)^2/(4D_2) - \gamma_1 xy/D_1$, such that the drift in \eqref{eq:model_asymmetric} is $\bm{R}=-{\rm diag}(D_1,D_2) \nabla F$.  This means that non-reciprocal interactions can be quantified by a single coupling constant $\gamma_1 D_2 + \gamma_2 D_1$.

\begin{figure}[t]
\includegraphics[width=\linewidth]{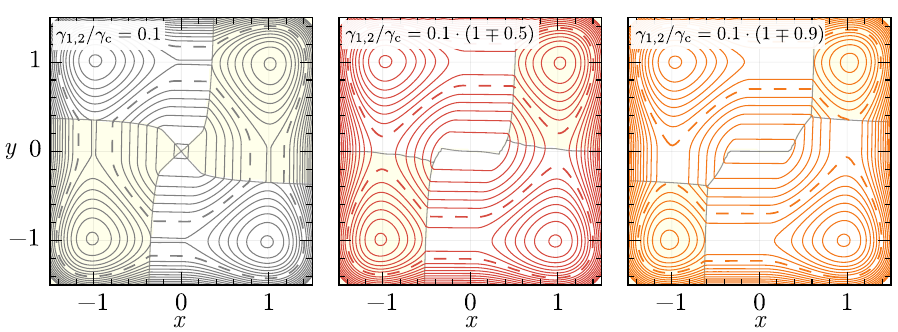}
\caption{
Global QP for asymmetric interactions $(\gamma_1, \gamma_2)$, computed numerically for increasing degrees of asymmetry. 
The left panel shows the symmetric case $\gamma_1 = \gamma_2$, which retains the $\mathbb{Z}_4$ symmetry discussed in previous sections. 
In contrast, the centre and right panels show asymmetric cases with $\gamma_1 < \gamma_2$, where the symmetry is reduced to $\mathbb{Z}_2$. 
This reduction is illustrated by the yellow shaded regions, which indicate domains where $F_\B$ or $F_\D$ locally minimise Eq.~\eqref{eq:global_from_local_qp}. 
Gray lines mark the switching lines $\Sigma_{\mathrm{bs}}$ where the minimising RQP in Eq.~\eqref{eq:global_from_local_qp} changes. 
To help visualise the asymmetry of the QP, the 7-th contour line is highlighted in a dashed style. 
Diffusion constants are set to $D_1 = D_2 = 1$.
}
\label{fig:qp_global_asym}
\end{figure}

Illustrative numerical results for this case are shown in Figure~\ref{fig:qp_global_asym}. 
 We take $D_1=D_2=1$ and parameterise $\gamma_{1,2} = \gamma(1\mp\lambda)$ so that $\lambda=0$ recovers $\mathbb Z_4$ symmetry.  Fixing $\gamma=0.1\gamma_c$, we increase $\lambda$: the reduction to $\mathbb Z_2$ symmetry is apparent because the four minima in the quasipotential landscape are no longer equally deep, and the basins surrounding them have different sizes. 
 In plotting these illustrative results we used the perturbative approximation of Eq.~\eqref{eq:offset-asym} below, which is sufficiently accurate for these purposes.  It reduces to $W_\A=W_\C=0$ and $W_\B=W_\D\approx 2\gamma \lambda$ for this case.
 
 A perturbative calculation of the RQPs for this reduced symmetry model is presented in App.~\ref{app:asymmetric_qp}, which also shows the accuracy of perturbation theory for the state points considered in Figure~\ref{fig:qp_global_asym}.  The zeroth order case has $\gamma_1=\gamma_2=0$ but $D_1\neq D_2$ in general, and we expand in  $\gamma_1,\gamma_2$.
For the Escher cycle, the most relevant quantities are the values at the saddles, which are analogous to $\Delta F^+$ from Eq.~\eqref{eq:F_1dex}; these set the barriers for transitions around the cycle {(see however~\cite{borner2024saddle})}.  
The reduced symmetry now means that barrier crossings from  $\B \to \C$ and $\D \to \A$ (``vertical'' in Figure~\ref{fig:qp_global_asym}) have a different barrier height from those from $\A \to \B$ and $\C \to \D$ (``horizontal'').  The perturbative calculation of the RQP yields its value at the saddle points:
\begin{align}\label{eq:barrier_height_asym}
\begin{split}
  F_\A(\ve{z}_{\sP}) &= F_\C(\ve{z}_{\sR}) = \frac1{D_1} \left(\frac14 - \gamma_1\right) + \ogg , 
  \\
  F_\B(\ve{z}_{\sQ}) &= F_\D(\ve{z}_{\sS}) = \frac1{D_2} \left(\frac14 - \gamma_2\right) + \ogg  . 
\end{split}
\end{align}

For the global quasipotential we use Eq.~\eqref{eq:global_from_local_qp}, which requires the basin offsets.  The $\mathbb{Z}_2$ symmetry means that $W_\A=W_\C$ and $W_\B=W_\D$.  As stated in Sec.~\ref{sec:glob-defns} these can be obtained from a Markov chain whose transition rates are $k_{X\to Y}={\rm e}^{-F_X(\z_Y)/\epsilon}$ which in the current system means 
\begin{align}
k_{\A\to\B} = k_{\C\to\D} \approx e^{-(1-4\gamma_1)/(4\epsilon D_1)} \qquad & k_{\B\to\A} = k_{\D\to\C} \approx e^{-(1+4\gamma_1)/(4\epsilon D_1)} 
\nonumber\\
k_{\B\to\C} = k_{\D\to\A} \approx e^{-(1-4\gamma_2)/(4\epsilon D_2)} \qquad & k_{\C\to\B} = k_{\A\to\D} \approx e^{-(1+4\gamma_2)/(4\epsilon D_2)} 
\end{align}
where the approximate equalities reflect that the arguments of the exponentials have been evaluated at first order in perturbation theory.  This Markov chain has steady state probabilities
\begin{equation}
\pi_\A = \pi_\C = \frac{ k_{\B\to\A} +  k_{\D\to\A} }{ 2(k_{\B\to\A} +  k_{\D\to\A} +k_{\A\to\B} +  k_{\C\to\B}) }  , \qquad 
\pi_\B = \pi_\D = \frac{ k_{\A\to\B} +  k_{\C\to\B} }{ 2(k_{\B\to\A} +  k_{\D\to\A} +k_{\A\to\B} +  k_{\C\to\B}) }
\end{equation}
Using $W_X=\lim_{\epsilon\to0} [-\epsilon \log \pi_X]$ yields
\begin{align}
\label{eq:offset-asym}
  W_\A=W_\C = \max \{ 0, -\Delta W \}
  , \qquad 
  W_\B=W_\D = \max \{ 0, \Delta W\} 
\end{align}
with  
\begin{equation}
\Delta W 
\approx 
\begin{cases} 
2\gamma_2/D_2 \; ,
& D_1 < D_2 \; ,
\\
(\gamma_2 - \gamma_1)/D_1 \; ,
\qquad  & D_1=D_2 \; ,
\\
-2\gamma_1/D_1 \; ,
& D_1 > D_2 \; .
\end{cases}
\end{equation}
The approximate equality here reflects that the result is accurate up to corrections at $\ogg$.  [Note also, this result is perturbative in $\gamma_1,\gamma_2$ with fixed $D_1,D_2$ so it does not cover cases with $D_1-D_2=\bigO(\gamma)$.]  As expected, larger $\gamma_1$ increases $\pi_\B,\pi_\D$ at the expense of $\pi_\A,\pi_\C$, while large $\gamma_2$ has the opposite effect.  

The results of Figure~\ref{fig:qp_global_asym} have $\gamma_1<\gamma_2$, hence the basins for $\B,\D$ are smaller (highlighted with the yellow shading). One also sees that the basins surrounding $\A,\C$ acquire a mutual boundary (grey line segment passing through the origin).  Physically, the main points to emphasize are the reduction from $\mathbb{Z}_4$ to $\mathbb{Z}_2$ symmetry, and the fact that the non-trivial basin offsets are computed from a Markov chain that describes the dynamics of hopping among the fixed points.

\section{Conclusion}\label{sec:conclusion}

We have analysed a model for two coupled {global} order parameters $x,y$ (or equivalently, a single particle moving in two dimensions).  We considered a low-noise limit where the dynamics is described by rare hopping events between fixed points of the noise-free dynamics, which is described by Freidlin--Wentzell theory~\cite{freidlin2012random}.  This allows computation of quasipotentials relative to each fixed point (RQPs), which are patched together to obtain the (global) quasipotential.  In general, this patching also requires computation of basin offsets, based on a Markov chain that describes hopping among the local minima of the quasipotential.

The model has a non-reciprocal coupling $\gamma$ between the order parameters -- it can be motivated as a mean-field version of the nonreciprocal Ising model of Avni \emph{et al.}~\cite{avni2023non,avni2024dynamical}. 
In the general case, this Markov chain has a steady state that breaks time-reversal symmetry, with rare transitions among its four basins, around the Escher cycle A,B,C,D.  Each step in the cycle occurs at an exponentially-distributed time whose mean scales as ${\rm e}^{-\Delta F/\epsilon}$ for an appropriate barrier (quasipotential difference) $\Delta F$.  Hence, the time taken to complete a Escher cycle is stochastic, in contrast to the limit cycles which occur in this model for larger $\gamma$, and have a fixed period.

For small values of $\gamma$, we analysed this behaviour in two ways: first by projecting the configuration space onto a one-dimensional reaction coordinate, and then by a perturbative calculation of the quasipotential as a function of $(x,y)$.  The perturbative calculation uses that the quasipotential solves a Hamilton-Jacobi equation, but it also uses a variational characterisation of the RQP in terms of instantons, see also \cite{graham1985weak,graham1986nonequilibrium}.  The analytical results rely on the simple structure of the model, but we argue that the qualitative behaviour is generic.   We also evaluate the RQP numerically using the method of Kikuchi \emph{et al}.~\cite{kikuchi2020ritz}: the results agree with perturbation theory for small $\gamma$ (as they should); the same qualitative behaviour is present for large $\gamma$ too.

The resulting QP landscapes display rich nonequilibrium signatures, combining smooth wells coexisting with flat plateaux and sharp, non-differentiable switching lines. The plateaux are the $2d$ continuation of $1d$ linear segments and arise from fluctuation pathways that pass through saddles; geometrically, the flat directions coincide with downhill parts leaving the saddle, along which the action does not increase. The non-differentiable features have two distinct origins. First, within a basin, multiple local minimising paths of the action can exist, and when their relative action cross, the global minimiser switches and produces a cusp in the RQP. In contrast to $1d$ where such non-differentiable sets are limited to points, we find non-differentiable lines in $2d$. We make the origin of such cusps explicit by stitching the actions along the competing smooth minimisers via a pointwise minimisation. Second, when assembling the global QP from local pieces across basins, the globally minimising RQP can also switch which produces additional kinks where the dominant RQP changes.

In a broader context, it may be possible to link the construction discussed here to nucleation in the full, finite nonreciprocal Ising model: each inter-basin transition in our surrogate corresponds to a ``zero-dimensional nucleation’’ event. If nucleation in the full model concentrates along a low-dimensional reaction coordinate, our barrier formulas should predict, or at least closely approximate, the associated rates; regardless, the structures we identify may still provide guidance for discovering such coordinates. In this sense, the present results both supply a minimal working example for nonequilibrium QPs and chart a practical route toward a nucleation theory beyond detailed balance, in systems where nonreciprocity is not a complication to be averaged away but the organising principle that shapes pathways and rates.

\section*{Acknowledgements}

We thank Cesare Nardini and Yael Avni for insightful discussions.
J.S. acknowledges funding through the UK Engineering and Physical Sciences Research Council (EPSRC) through grant number 2602536.  M.E.C. and R.L.J. also thank EPSRC for support through grant EP/Z534766/1.

\appendix

\section{Numerical calculation of quasipotential}
\label{app:numerics}

Sec.~\ref{sec:ldp_qp} outlined a direct strategy to compute the RQP: for each target point $\z$, determine the instanton connecting it to a reference basin by minimising the action, and evaluate the corresponding cost. While this procedure is generally not tractable analytically, it can be implemented numerically.
Several approaches exist for this minimisation, including the Minimum Action Method (MAM), string methods, and geometric MAM, which represent the path either in physical space or via arclength and iteratively relax it to lower the action \cite{weinan2004minimum,heymann2008geometric,grafke2019numerical,grafke2017long,heymann2008pathways,kikuchi2020ritz}. These algorithms have been successfully applied across a wide range of systems, from non-equilibrium Langevin models to active field theories \cite{zakine2023minimum}.

In this paper we employ the method of Kikuchi \emph{et al.}~\cite{kikuchi2020ritz}, which represents the instanton $\varphi$ in a Chebyshev polynomial basis, we take $n=10$ basis functions throughout. The action is evaluated along this path using Clenshaw--Curtis quadrature of order $10n$, and the coefficients are optimised with the NLopt library. This approach achieves high accuracy with relatively few parameters and avoids explicit time discretisation, making it efficient for instantons with infinite temporal support.
The configuration space $(x,y)$ is discretised on a grid $[-1.5,1.5]^2$ with $101$ points per dimension. We compute the RQP relative to $X=\A$ and obtain the others via symmetry.

The method of~\cite{kikuchi2020ritz} relies on a variational search, and appropriate
initialisation is essential to avoid convergence to spurious local minima. This is especially relevant at large $\gamma$, where indirect routes such as $\A \to \B \to \C \to \D$ may dominate over the direct $\A \to \D$ path. To capture this competition, we initialise five candidate paths for each $\z$:
(i) direct interpolation (i.e. $\A \to \z$),
(ii) interpolation via $\sP$ (i.e. $\A \to \sP \to \z$),
(iii) via $\B$ and $\sQ$ (i.e. $\A \to \B \to \sQ \to \z$),
(iv) via $\B$, $\C$, and $\D$ (i.e. $\A \to \B \to \C \to \D \to \z$),
(v) via $\sS$ (i.e. $\A \to \sS \to \z$).
This reduced set was selected from an initial pool of 10 candidate initial interpolations by coarse-grid screening. To each path we add small Gaussian noise. For every initial path we compute the optimised instanton for each initialisation, evaluate the corresponding action, and retain the minimal value. Although this procedure is computationally expensive, it robustly identifies the relevant instantons.

We repeat the above for every $\z$ in the configuration space. The outcome is a discretised map $\z \mapsto F_\A(\z)$. The global quasipotential $F(\z)$ is then assembled using Eq.~\eqref{eq:global_from_local_qp}.

\section{Perturbation theory}\label{app:failure}

This appendix discusses perturbative expansion of the quasipotential.  We use simple one-dimensional examples to illustrate issues that are relevant for the $2d$ model considered in main text.

\subsection{Perturbative local quasipotentials in an equilibrium system}
\label{app:eqm-pt}

As a very simple example consider $x\in\mathbb{R}$ with
\begin{equation}
\dot x = x - x^3 + \gamma + \sqrt{2\epsilon} \, \eta
\label{eq:sde_xinR}
\end{equation}
The global quasipotential is $U_\gamma(x)= \frac14 (1- x^2)^2 -\gamma x +C(\gamma)$ where the constant $C$ is chosen such that $U_\gamma(x^*)=0$ at its (global) minimum $x^*$.  This $U_\gamma$ has simple behaviour, including for perturbation theory around $\gamma=0$.  However, perturbative expansion of the local quasipotential is non-trivial, as we now discuss.

We take $|\gamma|<\sqrt{8}$, means that $U_\gamma$ has two local minima, the left one of which is at $x=x_\A$ [recall Eq.~\eqref{eq:x_A}]; its local maximum is at $x_P$.
The local quasipotential for basin $A$ is 
\begin{equation}
F_\A(x) = \begin{cases} 
-\gamma x + (x^2-1)^2/4 + C_\A(\gamma) , \qquad & x\leq x_P
\\
\Delta F(\gamma) & x > x_P
\end{cases} 
\label{equ:FA-trivial}
\end{equation}
where $C_\A(\gamma)= \frac14 (1- x_\A^2)^2 -\gamma x_\A$ so $F_\A(x_\A)=0$, and also $\Delta F(\gamma) = \frac14 (1- x_P^2)^2 - \gamma x_P + C_\A(\gamma)$, so $F_\A$ is continuous.

Perturbation theory for the quasipotential corresponds to an expansion in powers of $\gamma$ at fixed $x$ (with $x\neq 0$).  Expanding Eq.~\eqref{equ:FA-trivial} in powers $\gamma$ gives the perturbative expansion 
\begin{equation}
F_\A(x)  \approx F_\A^{\rm pert}(x) :=  \begin{cases} 
 \frac14 (1- x^2)^2 -\gamma (x+1)  \qquad & x< 0 \\
\frac14 - \gamma & x>0 
\end{cases}
\end{equation}
where the approximate equality indicates that we dropped terms at $O(\gamma^2)$.  [This result can also be obtained by the perturbative method discussed in Sec.~\ref{sec:perturbative}, as may be seen by integrating Eq.~\eqref{eq:perturbation} along a zeroth order local minimiser $\varphi^{(0)}$ with constant $y=1$.] 

\begin{figure}[t]
\includegraphics[width=0.75\linewidth]{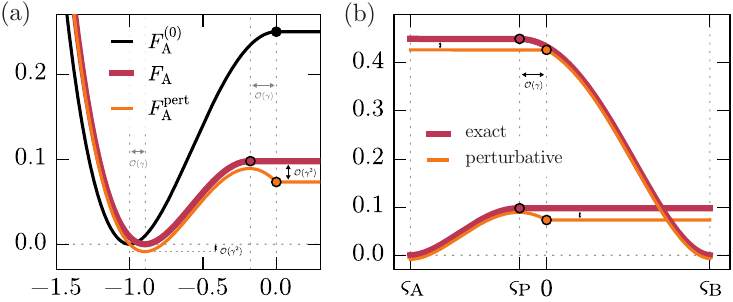}
\caption{Limitations and unphysical features of the perturbative expansion, for two cases discussed:
\textbf{(a)} Perturbative approximation to $F_\A$ in example of \eqref{eq:sde_xinR}, and 
\textbf{(b)} perturbative approximation to $F_\A$ and $F_\B$ in $1d$ reduced model, see Eqs.~\eqref{eq:F_A_pert_1d} and \eqref{eq:F_B_pert_1d}.
Both panels make obvious the effect of $\mathcal{O}(\gamma^2)$ deviation in the perturbative approximation (orange) compared to the exact expression (red), which demonstrate that the perturbative approximation can be negative and monotonically decreasing, both of which are unphysical for a local quasipotential. 
For illustrative purposes we show relatively large $\gamma = 0.5\gammac$. 
}
\label{fig:perturbation_failure}
\end{figure}

Fig.~\ref{fig:perturbation_failure}(a) shows that for small positive $\gamma$, the perturbative result $F_\A^{\rm pert}(x)$ is mostly close to $F_\A$,  but it has the wrong qualitative behaviour close to $x=0$, where it develops an unphysical local maximum.  We also note that continuing the perturbation expansion to higher order does not improve agreement between approximated and exact quasipotential in this region.  Indeed, $F_\A^{\rm pert}$ always has a singular point at $x=0$ while the true quasipotential is singular at $x=x_P$.  This reflects that the radius of convergence of the perturbation expansion is shrinking to zero as $x\to0$ so for any fixed $\gamma\neq 0$ there is a region close to $x=0$ where the perturbation theory does not converge.

These issues are generic for perturbation expansions of singular functions, so they should not be surprising.  However, the specific failure mode of this example -- where $F_\A^{\rm pert}(x)$ shows an unphysical maximum in a small interval of $x$ near a stationary point -- also appears in the $2d$ model.  We have demonstrated this effect in a $1d$ equilibrium model where the global quasipotential is simple; the local quasipotentials can still feature singularities, so perturbation theory fails.

\subsection{Perturbative approach to the one-dimensional reduced model}

In addition to the previous example, we also briefly describe how the perturbation theory of Sec.~\ref{sec:perturbative} applies to the reduced $1d$ model of Sec.~\ref{sec:1dqp}.  In particular, we show how the behaviour of the local quasipotentials $F_\A,F_\B$ near the saddle point $P$ is affected by the failure of perturbation theory.

For $\zo_\A\leq \zo \leq \zo_\B$ the analysis of the previous section applies.  At zeroth order we have $u_\A=-1$ and $u_\B=1$ so we obtain
\begin{equation}\label{eq:F_A_pert_1d}
F_\A^{\rm pert}(\zo) :=  \begin{cases} 
 (\zo^2-1)^2/4 -\gamma (\zo+1)  \qquad & -1 < \zo < 0 \\
(1/4) - \gamma & 0 < \zo < 1 
\end{cases}
\end{equation}
and
\begin{equation}\label{eq:F_B_pert_1d}
F_\B^{\rm pert}(\zo) :=  \begin{cases} 
   (1/4) - \gamma & -1 < \zo < 0 \\
 (\zo^2-1)^2/4 -\gamma (\zo-1) \qquad & 0 < \zo < 1 
\end{cases}
\end{equation}

Note also that for $\zo_\A\leq \zo \leq \zo_\B$ the global quasipotential is $F(\zo) = \min\{ F_\A(\zo),F_\B(\zo) \}$.  [This relies on $W_\A=W_\B$ in Eq.~(\ref{eq:global_from_local_qp}) and is different from the equilibrium case of App.~\ref{app:eqm-pt} where $W_\A-W_\B$ would be the difference between the two minima of $U_\gamma$.]

Fig.~\ref{fig:perturbation_failure}(b) shows the perturbative results together with the true quasipotentials.  
The spurious non-monotonicity of the perturbative quasipotential close to $x=0$ can lead to inaccurate predictions when comparing direct and indirect branches of the (local) quasipotential in the $2d$ case.  Appendix~\ref{app:branch_to_local} discusses an ad hoc method for estimating $F_\A(x,y)$ in that case, based on the perturbative calculation of Sec~\ref{sec:main}.  Roughly speaking, we obtain an approximate $F_\A^{\rm app}(x,y)$ from $F_\A^{\rm pert}(x,y)$ by adding a constant such that $\min[F_\A^{\rm pert}(x,y)]=0$ exactly, and removing spurious maxima by taking $F_\A^{\rm app}(x,y)$ constant in $x$ everywhere to the right of any local maximum.

\section{Patching of perturbative expansions for RQP branches}\label{app:branch_to_local}

In this section, we describe how we obtained plots of the RQP in Figs.~\ref{fig:qp_branches}, \ref{fig:qp_local} and  \ref{fig:qp_local_full}. This also feeds into the perturbative results shown in Fig.~\ref{fig:qp_global}, and also to Fig.~\ref{fig:qp_local_asym}.  This is achieved by Eq.~\eqref{eq:Fm_from_stationarypaths} but some care is needed to
overcome the limited applicability of perturbation theory described in App.~\ref{app:failure}, which becomes apparent through its nonmonotonic prediction for the RQP [Fig.~\ref{fig:perturbation_failure}].

Motivated by the results of App.~\ref{app:failure}, we 
modify the perturbative formula for the direct branch $ F^\rightarrow_\A $, computed via Eq.~\eqref{eq:F_A_branch1}, by enforcing that $(\partial F^\rightarrow_\A /\partial x) \geq 0$ for all $y\geq0$. 
Specifically, for each $y$-value, we search for the maximum of $ F^\rightarrow_\A $ along the $x$-direction within the interval $-0.75 \leq x \leq 0$, i.e. near the $y = 0$ line. The cutoff at $x = -0.75$ is chosen to lie safely between the stable point $ x_\A = -1 + \bigO(\gamma) $ and the saddle $ x_\sP = \bigO(\gamma) $.
If a local maximum is found at $x_{\rm max}$, we flatten the profile to the right of $x_{\rm max}$ by setting $ F^\rightarrow_\A(x, y) = F^\rightarrow_\A(x_{\rm max}, y) $ for all $x_{\rm max} \leq x \leq 0$
[this is the region highlighted in red in Fig.~\ref{fig:qp_branches}(a)]. This ensures that the adjusted $ F^\rightarrow_\A $ is non-decreasing in $x$.  It is this adjusted $ F^\rightarrow_\A $ that is used in Eq.~\eqref{eq:Fm_from_stationarypaths} for numerical estimation of $F_\A$.  Comparing with Fig.~\ref{fig:perturbation_failure} one sees that the adjustment to $ F^\rightarrow_\A $ scales as $\bigO(\gamma^2)$, so the result is still accurate to first order.

\section{Reduced symmetry case: Perturbation theory for the RQP}
\label{app:asymmetric_qp}

In this appendix, we discuss in more detail the reduced symmetry case introduced in Sec.~\ref{sec:asymmetric}. 
The idea is to extend the model in Eq.~\eqref{eq:model} to a case in which the quasipotential reduces its fourfold degeneracy of the stable fixed points to two pairwise-degenerate basins. To achieve this, we generalised the model to include asymmetry in both the nonreciprocal coupling and noise amplitudes. As described in the main text, we introduce two independent interaction strengths, $\gamma_1$ and $\gamma_2$, and two independent diffusion coefficients $D_1$ and $D_2$. This leads to the generalised model in Eqs.~\eqref{eq:model_asymmetric}.
It is helpful to reparametrise the interaction strengths as
\begin{align}\label{eq:gammas_split}
  \gamma_1 = \gamma(1 + \lambda), \quad \gamma_2 = \gamma(1 - \lambda),
\end{align}
where $\gamma$ is a small parameter and $\lambda \in [-1,1]$ quantifies the degree of ``asymmetry''. For $\lambda = 0$, we recover the original model studied in Eq.~\eqref{eq:model}.
In the parametrisation with $\gamma$ and $\lambda$, the  model in Eqs.~\eqref{eq:model_asymmetric} becomes
\begin{align}\label{eq:model_asymmetric2} 
\begin{split}
\dot{x} &= x - x^3 + \gamma \lambda y + \gamma y + \sqrt{2\epsilon D_1} \, \eta_1(t) ,
\\
\dot{y} &= y - y^3 + \gamma \lambda x - \gamma x + \sqrt{2\epsilon D_2} \, \eta_2(t) .
\end{split}
\end{align}

The system no longer preserves full $\mathbb{Z}_4$ rotational symmetry in the configuration space but retains a residual $180^\circ$ ($\mathbb{Z}_2$) rotational symmetry. 
However, the system remains symmetric under a $90^\circ$ rotation combined with the exchange $D_1 \leftrightarrow D_2$ and $\gamma_1 \leftrightarrow \gamma_2$  (the latter is equivalent to $\lambda \to -\lambda$). This helps mapping between RQPs for different fixed points.
We can observe this combined symmetry in the expansion of the fixed points:
for the original model we found an expansion of the fixed points in  Eqs.~\eqref{eq:fp_expansion}, while for the generalised model we have
\begin{subequations}\label{eq:stable_fp_full_expansion_asym}
\begin{align}
  &x_{\A} = -1 + \frac{1+\lambda}2 \gamma + \frac{(1+\lambda)(5+\lambda)}8 \gamma^2 + \bigO(\gamma^3),
  \qquad 
  &&y_{\A} = +1 + \frac{1-\lambda}2 \gamma - \frac{(1-\lambda)(5-\lambda)}8 \gamma^2+ \bigO(\gamma^3),
  \\
  &x_{\sP} = - (1+\lambda)\gamma + \bigO(\gamma^3),
  \qquad 
  &&y_{\sP} = +1 + \frac{1-\lambda^2}2 \gamma^2 + \bigO(\gamma^3),
  \\
  &x_{\B} = +1 + \frac{1+\lambda}2 \gamma - \frac{(1+\lambda)(5+\lambda)}8 \gamma^2 + \bigO(\gamma^3),
  \qquad 
  &&y_{\B} = +1 - \frac{1-\lambda}2 \gamma - \frac{(1-\lambda)(5-\lambda)}8 \gamma^2 + \bigO(\gamma^3).
\end{align}
\end{subequations}
The broken symmetry means that $\ve{z}_\B$ cannot be obtained from $\ve{z}_\A$ by a simple $90^\circ$ rotation, but requires the additional transformation $\lambda \to -\lambda$.

\begin{figure}[t]
\includegraphics[width=\linewidth]{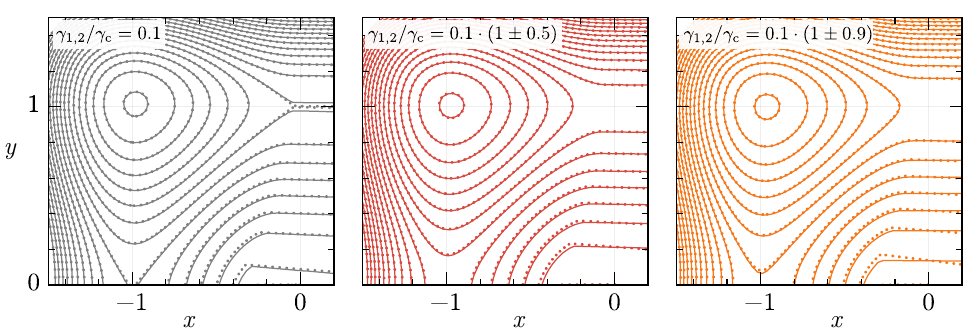}
\caption{
RQP $F_\A$ in the reduced symmetry case with interactions $\gamma_{1}, \gamma_2$ for increasing $\lambda$.
Lines show numerical evaluation, dots show analytical expressions from Eq.~\eqref{eq:F_A_generalized}. 
Diffusion constants are set to $D_1 = D_2 = 1$. 
}
\label{fig:qp_local_asym}
\end{figure}

We now derive the branches of the action in the reduced symmetry model [Eqs.~\eqref{eq:model_asymmetric2}].  The zeroth order in perturbation theory is obtained by setting $\gamma=0$, so the zeroth order RQP
$F_X^{(0)}$ and the associated instantons are the same as in Sec.~\ref{sec:branches}, aside from the additional factors $D_1$ and $D_2$, and appropriate replacement of $\gamma$ by $\gamma_1,\gamma_2$, as appropriate.
Using this with Eq.~\eqref{eq:perturbation}, 
the calculation of the direct branch of the RQP from $\A$ proceeds analogously to Sec.~\ref{sec:direct_branch}, yielding
\begin{align}
    F_{\A}^\rightarrow(\ve{z}) = \frac{1}{4D_1}(1 - x^2)^2 + \frac{1}{4D_2}(1 - y^2)^2 + \frac{\gamma_1}{D_1} \frac{(1-x^2)y}{x - y} + \frac{\gamma_2}{D_2} \frac{(1-y^2)x}{x - y} +\ogg 
\end{align}
Applying the combined $90^\circ$ rotation and parameter swap ($\gamma_1\leftrightarrow\gamma_2$ and $D_1\leftrightarrow D_2$) we obtain the corresponding branch from $\B$
\begin{align}
    F_{\B}^\rightarrow(\ve{z}) = \frac{1}{4D_1}(1 - x^2)^2 + \frac{1}{4D_2}(1 - y^2)^2 
    - \frac{\gamma_1}{D_1} \frac{(1-x^2)y}{ x + y }
    + \frac{\gamma_2}{D_2} \frac{(1-y^2)x}{ x + y }  
    + \ogg 
\end{align}
This leads to the barrier heights of Eq.~\eqref{eq:barrier_height_asym}, via
\begin{align}\label{eq:barrier_height_asym_app}
\begin{split}
  F^\rightarrow_\A(\ve{z}_\sP) = \frac1{D_1} \left(\frac14 - \gamma_1\right) +\ogg 
  \\
  F^\rightarrow_\B(\ve{z}_\sQ)  = \frac1{D_2} \left(\frac14 - \gamma_2\right) +\ogg 
\end{split}
\end{align}

Similarly, following  Sec.~\ref{sec:indirect_branch}, we find the corresponding RQPs relative to the saddles as 
\begin{align}\label{eq:direct_generalized_saddle}
  F_{\sP}^{\rightarrow}(\ve{z})= \frac{1}{4D_2}(1 - y^2)^2 - \frac{\gamma_2}{D_2} xy(1-y^2) \int_0^1 \frac{u^2 \plaind{u}}{[(1-x^2)+x^2u^2]^{1/2} \, [y^2 + (1-y^2)u^2]^{3/2}} + \ogg 
  \\
  F_{\sQ}^{\rightarrow}(\ve{z})= \frac{1}{4D_1}(1 - x^2)^2 + \frac{\gamma_1}{D_1} xy(1-x^2) \int_0^1 \frac{u^2 \plaind{u}}{[(1-y^2)+y^2u^2]^{1/2} \, [x^2 + (1-x^2)u^2]^{3/2}} + \ogg 
\end{align}
Combining these with the barriers  in Eq.~\eqref{eq:barrier_height_asym_app} and using Eq.~\eqref{eq:F_A_indirect} yields the indirect RQPs $F^\curvearrowright_\A$ and $F^\curvearrowright_\B$ for the generalised model.
Finally, the RQPs $F_\A$ and $F_\B$ are constructed from the direct and indirect branches in the same way as outlined in Sec.~\ref{sec:local_qp}, 
\begin{equation}\label{eq:F_A_generalized}
  F_\A(\ve{z}) = \min\big\{ F_{\A}^\rightarrow(\ve{z}), F_{\A}^\curvearrowright(\ve{z}) \big\} .
\end{equation}
The resulting RQP $F_\A$ is shown in Fig.~\ref{fig:qp_local_asym}, for increasing values of $\lambda$.  The agreement between the perturbative and numerically exact calculations is excellent.

\bibliography{refs}

\begin{thebibliography}{48}%
\makeatletter
\providecommand \@ifxundefined [1]{%
 \@ifx{#1\undefined}
}%
\providecommand \@ifnum [1]{%
 \ifnum #1\expandafter \@firstoftwo
 \else \expandafter \@secondoftwo
 \fi
}%
\providecommand \@ifx [1]{%
 \ifx #1\expandafter \@firstoftwo
 \else \expandafter \@secondoftwo
 \fi
}%
\providecommand \natexlab [1]{#1}%
\providecommand \enquote  [1]{``#1''}%
\providecommand \bibnamefont  [1]{#1}%
\providecommand \bibfnamefont [1]{#1}%
\providecommand \citenamefont [1]{#1}%
\providecommand \href@noop [0]{\@secondoftwo}%
\providecommand \href [0]{\begingroup \@sanitize@url \@href}%
\providecommand \@href[1]{\@@startlink{#1}\@@href}%
\providecommand \@@href[1]{\endgroup#1\@@endlink}%
\providecommand \@sanitize@url [0]{\catcode `\\12\catcode `\$12\catcode
  `\&12\catcode `\#12\catcode `\^12\catcode `\_12\catcode `\%12\relax}%
\providecommand \@@startlink[1]{}%
\providecommand \@@endlink[0]{}%
\providecommand \url  [0]{\begingroup\@sanitize@url \@url }%
\providecommand \@url [1]{\endgroup\@href {#1}{\urlprefix }}%
\providecommand \urlprefix  [0]{URL }%
\providecommand \Eprint [0]{\href }%
\providecommand \doibase [0]{https://doi.org/}%
\providecommand \selectlanguage [0]{\@gobble}%
\providecommand \bibinfo  [0]{\@secondoftwo}%
\providecommand \bibfield  [0]{\@secondoftwo}%
\providecommand \translation [1]{[#1]}%
\providecommand \BibitemOpen [0]{}%
\providecommand \bibitemStop [0]{}%
\providecommand \bibitemNoStop [0]{.\EOS\space}%
\providecommand \EOS [0]{\spacefactor3000\relax}%
\providecommand \BibitemShut  [1]{\csname bibitem#1\endcsname}%
\let\auto@bib@innerbib\@empty
\bibitem [{\citenamefont {O'Loan}\ and\ \citenamefont
  {Evans}(1999)}]{oloan1999alternating}%
  \BibitemOpen
  \bibfield  {author} {\bibinfo {author} {\bibfnamefont {O.~J.}\ \bibnamefont
  {O'Loan}}\ and\ \bibinfo {author} {\bibfnamefont {M.~R.}\ \bibnamefont
  {Evans}},\ }\bibfield  {title} {\bibinfo {title} {Alternating steady state in
  one-dimensional flocking},\ }\href
  {https://doi.org/10.1088/0305-4470/32/8/002} {\bibfield  {journal} {\bibinfo
  {journal} {J. Phys. A: Math. Gen.}\ }\textbf {\bibinfo {volume} {32}},\
  \bibinfo {pages} {L99} (\bibinfo {year} {1999})}\BibitemShut {NoStop}%
\bibitem [{\citenamefont {Besse}\ \emph {et~al.}(2022)\citenamefont {Besse},
  \citenamefont {Chat{\'e}},\ and\ \citenamefont
  {Solon}}]{besse2022metastability}%
  \BibitemOpen
  \bibfield  {author} {\bibinfo {author} {\bibfnamefont {M.}~\bibnamefont
  {Besse}}, \bibinfo {author} {\bibfnamefont {H.}~\bibnamefont {Chat{\'e}}},\
  and\ \bibinfo {author} {\bibfnamefont {A.}~\bibnamefont {Solon}},\ }\bibfield
   {title} {\bibinfo {title} {{Metastability of Constant-Density Flocks}},\
  }\href {https://doi.org/10.1103/physrevlett.129.268003} {\bibfield  {journal}
  {\bibinfo  {journal} {Phys. Rev. Lett.}\ }\textbf {\bibinfo {volume} {129}},\
  \bibinfo {pages} {268003} (\bibinfo {year} {2022})}\BibitemShut {NoStop}%
\bibitem [{\citenamefont {Benvegnen}\ \emph {et~al.}(2023)\citenamefont
  {Benvegnen}, \citenamefont {Granek}, \citenamefont {Ro}, \citenamefont
  {Yaacoby}, \citenamefont {Chat\'e}, \citenamefont {Kafri}, \citenamefont
  {Mukamel}, \citenamefont {Solon},\ and\ \citenamefont
  {Tailleur}}]{benvegnen2023metastability}%
  \BibitemOpen
  \bibfield  {author} {\bibinfo {author} {\bibfnamefont {B.}~\bibnamefont
  {Benvegnen}}, \bibinfo {author} {\bibfnamefont {O.}~\bibnamefont {Granek}},
  \bibinfo {author} {\bibfnamefont {S.}~\bibnamefont {Ro}}, \bibinfo {author}
  {\bibfnamefont {R.}~\bibnamefont {Yaacoby}}, \bibinfo {author} {\bibfnamefont
  {H.}~\bibnamefont {Chat\'e}}, \bibinfo {author} {\bibfnamefont
  {Y.}~\bibnamefont {Kafri}}, \bibinfo {author} {\bibfnamefont
  {D.}~\bibnamefont {Mukamel}}, \bibinfo {author} {\bibfnamefont
  {A.}~\bibnamefont {Solon}},\ and\ \bibinfo {author} {\bibfnamefont
  {J.}~\bibnamefont {Tailleur}},\ }\bibfield  {title} {\bibinfo {title}
  {{Metastability of Discrete-Symmetry Flocks}},\ }\href
  {https://doi.org/10.1103/physrevlett.131.218301} {\bibfield  {journal}
  {\bibinfo  {journal} {Phys. Rev. Lett.}\ }\textbf {\bibinfo {volume} {131}},\
  \bibinfo {pages} {218301} (\bibinfo {year} {2023})}\BibitemShut {NoStop}%
\bibitem [{\citenamefont {Grafke}\ \emph
  {et~al.}(2017{\natexlab{a}})\citenamefont {Grafke}, \citenamefont {Cates},\
  and\ \citenamefont {Vanden-Eijnden}}]{grafke2017spatiotemporal}%
  \BibitemOpen
  \bibfield  {author} {\bibinfo {author} {\bibfnamefont {T.}~\bibnamefont
  {Grafke}}, \bibinfo {author} {\bibfnamefont {M.~E.}\ \bibnamefont {Cates}},\
  and\ \bibinfo {author} {\bibfnamefont {E.}~\bibnamefont {Vanden-Eijnden}},\
  }\bibfield  {title} {\bibinfo {title} {{Spatiotemporal Self-Organization of
  Fluctuating Bacterial Colonies}},\ }\href
  {https://doi.org/10.1103/physrevlett.119.188003} {\bibfield  {journal}
  {\bibinfo  {journal} {Phys. Rev. Lett.}\ }\textbf {\bibinfo {volume} {119}},\
  \bibinfo {pages} {188003} (\bibinfo {year} {2017}{\natexlab{a}})}\BibitemShut
  {NoStop}%
\bibitem [{\citenamefont {Noguchi}\ and\ \citenamefont
  {Fournier}(2024)}]{noguchi2024Spatiotemporal}%
  \BibitemOpen
  \bibfield  {author} {\bibinfo {author} {\bibfnamefont {H.}~\bibnamefont
  {Noguchi}}\ and\ \bibinfo {author} {\bibfnamefont {J.-B.}\ \bibnamefont
  {Fournier}},\ }\bibfield  {title} {\bibinfo {title} {Spatiotemporal patterns
  in the active cyclic {P}otts model},\ }\href
  {https://doi.org/10.1088/1367-2630/ad7dac} {\bibfield  {journal} {\bibinfo
  {journal} {New J. Phys.}\ }\textbf {\bibinfo {volume} {26}},\ \bibinfo
  {pages} {093043} (\bibinfo {year} {2024})}\BibitemShut {NoStop}%
\bibitem [{\citenamefont {Noguchi}(2025)}]{noguchi2025spatiotemporal}%
  \BibitemOpen
  \bibfield  {author} {\bibinfo {author} {\bibfnamefont {H.}~\bibnamefont
  {Noguchi}},\ }\bibfield  {title} {\bibinfo {title} {{Spatiotemporal patterns
  in active four-state Potts models}},\ }\href
  {https://doi.org/10.1038/s41598-024-84819-w} {\bibfield  {journal} {\bibinfo
  {journal} {Sci. Rep.}\ }\textbf {\bibinfo {volume} {15}},\ \bibinfo {pages}
  {674} (\bibinfo {year} {2025})}\BibitemShut {NoStop}%
\bibitem [{\citenamefont {Avni}\ \emph
  {et~al.}(2025{\natexlab{a}})\citenamefont {Avni}, \citenamefont {Fruchart},
  \citenamefont {Martin}, \citenamefont {Seara},\ and\ \citenamefont
  {Vitelli}}]{avni2023non}%
  \BibitemOpen
  \bibfield  {author} {\bibinfo {author} {\bibfnamefont {Y.}~\bibnamefont
  {Avni}}, \bibinfo {author} {\bibfnamefont {M.}~\bibnamefont {Fruchart}},
  \bibinfo {author} {\bibfnamefont {D.}~\bibnamefont {Martin}}, \bibinfo
  {author} {\bibfnamefont {D.}~\bibnamefont {Seara}},\ and\ \bibinfo {author}
  {\bibfnamefont {V.}~\bibnamefont {Vitelli}},\ }\bibfield  {title} {\bibinfo
  {title} {{Nonreciprocal Ising Model}},\ }\href
  {https://doi.org/10.1103/physrevlett.134.117103} {\bibfield  {journal}
  {\bibinfo  {journal} {Phys. Rev. Lett.}\ }\textbf {\bibinfo {volume} {134}},\
  \bibinfo {pages} {117103} (\bibinfo {year} {2025}{\natexlab{a}})}\BibitemShut
  {NoStop}%
\bibitem [{\citenamefont {Avni}\ \emph
  {et~al.}(2025{\natexlab{b}})\citenamefont {Avni}, \citenamefont {Fruchart},
  \citenamefont {Martin}, \citenamefont {Seara},\ and\ \citenamefont
  {Vitelli}}]{avni2024dynamical}%
  \BibitemOpen
  \bibfield  {author} {\bibinfo {author} {\bibfnamefont {Y.}~\bibnamefont
  {Avni}}, \bibinfo {author} {\bibfnamefont {M.}~\bibnamefont {Fruchart}},
  \bibinfo {author} {\bibfnamefont {D.}~\bibnamefont {Martin}}, \bibinfo
  {author} {\bibfnamefont {D.}~\bibnamefont {Seara}},\ and\ \bibinfo {author}
  {\bibfnamefont {V.}~\bibnamefont {Vitelli}},\ }\bibfield  {title} {\bibinfo
  {title} {Dynamical phase transitions in the nonreciprocal {I}sing model},\
  }\href {https://doi.org/10.1103/physreve.111.034124} {\bibfield  {journal}
  {\bibinfo  {journal} {Phys. Rev. E}\ }\textbf {\bibinfo {volume} {111}},\
  \bibinfo {pages} {034124} (\bibinfo {year} {2025}{\natexlab{b}})}\BibitemShut
  {NoStop}%
\bibitem [{\citenamefont {You}\ \emph {et~al.}(2020)\citenamefont {You},
  \citenamefont {Baskaran},\ and\ \citenamefont
  {Marchetti}}]{you2020nonreciprocity}%
  \BibitemOpen
  \bibfield  {author} {\bibinfo {author} {\bibfnamefont {Z.}~\bibnamefont
  {You}}, \bibinfo {author} {\bibfnamefont {A.}~\bibnamefont {Baskaran}},\ and\
  \bibinfo {author} {\bibfnamefont {M.~C.}\ \bibnamefont {Marchetti}},\
  }\bibfield  {title} {\bibinfo {title} {Nonreciprocity as a generic route to
  traveling states},\ }\href {https://doi.org/10.1073/pnas.2010318117}
  {\bibfield  {journal} {\bibinfo  {journal} {Proc. Natl. Acad. Sci.}\ }\textbf
  {\bibinfo {volume} {117}},\ \bibinfo {pages} {19767} (\bibinfo {year}
  {2020})}\BibitemShut {NoStop}%
\bibitem [{\citenamefont {Fruchart}\ \emph {et~al.}(2021)\citenamefont
  {Fruchart}, \citenamefont {Hanai}, \citenamefont {Littlewood},\ and\
  \citenamefont {Vitelli}}]{fruchart2021non}%
  \BibitemOpen
  \bibfield  {author} {\bibinfo {author} {\bibfnamefont {M.}~\bibnamefont
  {Fruchart}}, \bibinfo {author} {\bibfnamefont {R.}~\bibnamefont {Hanai}},
  \bibinfo {author} {\bibfnamefont {P.~B.}\ \bibnamefont {Littlewood}},\ and\
  \bibinfo {author} {\bibfnamefont {V.}~\bibnamefont {Vitelli}},\ }\bibfield
  {title} {\bibinfo {title} {Non-reciprocal phase transitions},\ }\href
  {https://doi.org/10.1038/s41586-021-03375-9} {\bibfield  {journal} {\bibinfo
  {journal} {Nature}\ }\textbf {\bibinfo {volume} {592}},\ \bibinfo {pages}
  {363} (\bibinfo {year} {2021})}\BibitemShut {NoStop}%
\bibitem [{\citenamefont {Shankar}\ \emph {et~al.}(2022)\citenamefont
  {Shankar}, \citenamefont {Souslov}, \citenamefont {Bowick}, \citenamefont
  {Marchetti},\ and\ \citenamefont {Vitelli}}]{shankar2022topological}%
  \BibitemOpen
  \bibfield  {author} {\bibinfo {author} {\bibfnamefont {S.}~\bibnamefont
  {Shankar}}, \bibinfo {author} {\bibfnamefont {A.}~\bibnamefont {Souslov}},
  \bibinfo {author} {\bibfnamefont {M.~J.}\ \bibnamefont {Bowick}}, \bibinfo
  {author} {\bibfnamefont {M.~C.}\ \bibnamefont {Marchetti}},\ and\ \bibinfo
  {author} {\bibfnamefont {V.}~\bibnamefont {Vitelli}},\ }\bibfield  {title}
  {\bibinfo {title} {Topological active matter},\ }\href
  {https://doi.org/10.1038/s42254-022-00445-3} {\bibfield  {journal} {\bibinfo
  {journal} {Nat. Rev. Phys.}\ }\textbf {\bibinfo {volume} {4}},\ \bibinfo
  {pages} {380} (\bibinfo {year} {2022})}\BibitemShut {NoStop}%
\bibitem [{\citenamefont {Peters}(2017)}]{peters-book}%
  \BibitemOpen
  \bibfield  {author} {\bibinfo {author} {\bibfnamefont {B.}~\bibnamefont
  {Peters}},\ }\href@noop {} {\emph {\bibinfo {title} {Reaction Rate Theory and
  Rare Events Simulations}}}\ (\bibinfo  {publisher} {Elsevier},\ \bibinfo
  {year} {2017})\BibitemShut {NoStop}%
\bibitem [{\citenamefont {Cates}\ and\ \citenamefont
  {Nardini}(2023)}]{cates2023classical}%
  \BibitemOpen
  \bibfield  {author} {\bibinfo {author} {\bibfnamefont {M.~E.}\ \bibnamefont
  {Cates}}\ and\ \bibinfo {author} {\bibfnamefont {C.}~\bibnamefont
  {Nardini}},\ }\bibfield  {title} {\bibinfo {title} {{Classical Nucleation
  Theory for Active Fluid Phase Separation}},\ }\href
  {https://doi.org/10.1103/physrevlett.130.098203} {\bibfield  {journal}
  {\bibinfo  {journal} {Phys. Rev. Lett.}\ }\textbf {\bibinfo {volume} {130}},\
  \bibinfo {pages} {098203} (\bibinfo {year} {2023})}\BibitemShut {NoStop}%
\bibitem [{\citenamefont {Woillez}\ \emph {et~al.}(2019)\citenamefont
  {Woillez}, \citenamefont {Zhao}, \citenamefont {Kafri}, \citenamefont
  {Lecomte},\ and\ \citenamefont {Tailleur}}]{woillez2019activated}%
  \BibitemOpen
  \bibfield  {author} {\bibinfo {author} {\bibfnamefont {E.}~\bibnamefont
  {Woillez}}, \bibinfo {author} {\bibfnamefont {Y.}~\bibnamefont {Zhao}},
  \bibinfo {author} {\bibfnamefont {Y.}~\bibnamefont {Kafri}}, \bibinfo
  {author} {\bibfnamefont {V.}~\bibnamefont {Lecomte}},\ and\ \bibinfo {author}
  {\bibfnamefont {J.}~\bibnamefont {Tailleur}},\ }\bibfield  {title} {\bibinfo
  {title} {{Activated Escape of a Self-Propelled Particle from a Metastable
  State}},\ }\href {https://doi.org/10.1103/physrevlett.122.258001} {\bibfield
  {journal} {\bibinfo  {journal} {Phys. Rev. Lett.}\ }\textbf {\bibinfo
  {volume} {122}},\ \bibinfo {pages} {258001} (\bibinfo {year}
  {2019})}\BibitemShut {NoStop}%
\bibitem [{\citenamefont {Freidlin}\ and\ \citenamefont
  {Wentzell}(2012)}]{freidlin2012random}%
  \BibitemOpen
  \bibfield  {author} {\bibinfo {author} {\bibfnamefont {M.~I.}\ \bibnamefont
  {Freidlin}}\ and\ \bibinfo {author} {\bibfnamefont {A.~D.}\ \bibnamefont
  {Wentzell}},\ }\href {https://doi.org/10.1007/978-3-642-25847-3} {\emph
  {\bibinfo {title} {Random {{Perturbations}} of {{Dynamical Systems}}}}},\
  \bibinfo {series} {Grundlehren der mathematischen Wissenschaften}, Vol.\
  \bibinfo {volume} {260}\ (\bibinfo  {publisher} {Springer},\ \bibinfo
  {address} {Berlin, Heidelberg},\ \bibinfo {year} {2012})\BibitemShut
  {NoStop}%
\bibitem [{\citenamefont {Dykman}\ \emph {et~al.}(1994)\citenamefont {Dykman},
  \citenamefont {Mori}, \citenamefont {Ross},\ and\ \citenamefont
  {Hunt}}]{dykman1994large}%
  \BibitemOpen
  \bibfield  {author} {\bibinfo {author} {\bibfnamefont {M.~I.}\ \bibnamefont
  {Dykman}}, \bibinfo {author} {\bibfnamefont {E.}~\bibnamefont {Mori}},
  \bibinfo {author} {\bibfnamefont {J.}~\bibnamefont {Ross}},\ and\ \bibinfo
  {author} {\bibfnamefont {P.~M.}\ \bibnamefont {Hunt}},\ }\bibfield  {title}
  {\bibinfo {title} {Large fluctuations and optimal paths in chemical
  kinetics},\ }\href {https://doi.org/10.1063/1.467139} {\bibfield  {journal}
  {\bibinfo  {journal} {J. Chem. Phys.}\ }\textbf {\bibinfo {volume} {100}},\
  \bibinfo {pages} {5735} (\bibinfo {year} {1994})}\BibitemShut {NoStop}%
\bibitem [{\citenamefont {Lazarescu}\ \emph {et~al.}(2019)\citenamefont
  {Lazarescu}, \citenamefont {Cossetto}, \citenamefont {Falasco},\ and\
  \citenamefont {Esposito}}]{lazarescu2019large}%
  \BibitemOpen
  \bibfield  {author} {\bibinfo {author} {\bibfnamefont {A.}~\bibnamefont
  {Lazarescu}}, \bibinfo {author} {\bibfnamefont {T.}~\bibnamefont {Cossetto}},
  \bibinfo {author} {\bibfnamefont {G.}~\bibnamefont {Falasco}},\ and\ \bibinfo
  {author} {\bibfnamefont {M.}~\bibnamefont {Esposito}},\ }\bibfield  {title}
  {\bibinfo {title} {Large deviations and dynamical phase transitions in
  stochastic chemical networks},\ }\href {https://doi.org/10.1063/1.5111110}
  {\bibfield  {journal} {\bibinfo  {journal} {J. Chem. Phys.}\ }\textbf
  {\bibinfo {volume} {151}},\ \bibinfo {pages} {064117} (\bibinfo {year}
  {2019})}\BibitemShut {NoStop}%
\bibitem [{\citenamefont {Sch\"uttler}\ \emph {et~al.}(2024)\citenamefont
  {Sch\"uttler}, \citenamefont {Jack},\ and\ \citenamefont
  {Cates}}]{schuttler2024effects}%
  \BibitemOpen
  \bibfield  {author} {\bibinfo {author} {\bibfnamefont {J.}~\bibnamefont
  {Sch\"uttler}}, \bibinfo {author} {\bibfnamefont {R.~L.}\ \bibnamefont
  {Jack}},\ and\ \bibinfo {author} {\bibfnamefont {M.~E.}\ \bibnamefont
  {Cates}},\ }\bibfield  {title} {\bibinfo {title} {Effects of phase separation
  on extinction times in population models},\ }\href
  {https://doi.org/10.1088/1742-5468/ad5c59} {\bibfield  {journal} {\bibinfo
  {journal} {J. Stat. Mech.}\ }\textbf {\bibinfo {volume} {2024}},\ \bibinfo
  {pages} {083209} (\bibinfo {year} {2024})}\BibitemShut {NoStop}%
\bibitem [{\citenamefont {Heller}\ and\ \citenamefont
  {Limmer}(2024)}]{heller2024evaluation}%
  \BibitemOpen
  \bibfield  {author} {\bibinfo {author} {\bibfnamefont {E.~R.}\ \bibnamefont
  {Heller}}\ and\ \bibinfo {author} {\bibfnamefont {D.~T.}\ \bibnamefont
  {Limmer}},\ }\bibfield  {title} {\bibinfo {title} {Evaluation of transition
  rates from nonequilibrium instantons},\ }\href
  {https://doi.org/10.1103/PhysRevResearch.6.043110} {\bibfield  {journal}
  {\bibinfo  {journal} {Phys. Rev. Res.}\ }\textbf {\bibinfo {volume} {6}},\
  \bibinfo {pages} {043110} (\bibinfo {year} {2024})}\BibitemShut {NoStop}%
\bibitem [{\citenamefont {Crisanti}\ and\ \citenamefont
  {Paoluzzi}(2023)}]{crisanti2023most}%
  \BibitemOpen
  \bibfield  {author} {\bibinfo {author} {\bibfnamefont {A.}~\bibnamefont
  {Crisanti}}\ and\ \bibinfo {author} {\bibfnamefont {M.}~\bibnamefont
  {Paoluzzi}},\ }\bibfield  {title} {\bibinfo {title} {Most probable path of
  active {O}rnstein--{U}hlenbeck particles},\ }\href
  {https://doi.org/10.1103/PhysRevE.107.034110} {\bibfield  {journal} {\bibinfo
   {journal} {Phys. Rev. E}\ }\textbf {\bibinfo {volume} {107}},\ \bibinfo
  {pages} {034110} (\bibinfo {year} {2023})}\BibitemShut {NoStop}%
\bibitem [{\citenamefont {Yasuda}\ and\ \citenamefont
  {Ishimoto}(2022)}]{yasuda2022most}%
  \BibitemOpen
  \bibfield  {author} {\bibinfo {author} {\bibfnamefont {K.}~\bibnamefont
  {Yasuda}}\ and\ \bibinfo {author} {\bibfnamefont {K.}~\bibnamefont
  {Ishimoto}},\ }\bibfield  {title} {\bibinfo {title} {Most probable path of an
  active {B}rownian particle},\ }\href
  {https://doi.org/10.1103/PhysRevE.106.064120} {\bibfield  {journal} {\bibinfo
   {journal} {Phys. Rev. E}\ }\textbf {\bibinfo {volume} {106}},\ \bibinfo
  {pages} {064120} (\bibinfo {year} {2022})}\BibitemShut {NoStop}%
\bibitem [{\citenamefont {Goswami}\ and\ \citenamefont
  {Metzler}(2023)}]{goswami2023effects}%
  \BibitemOpen
  \bibfield  {author} {\bibinfo {author} {\bibfnamefont {K.}~\bibnamefont
  {Goswami}}\ and\ \bibinfo {author} {\bibfnamefont {R.}~\bibnamefont
  {Metzler}},\ }\bibfield  {title} {\bibinfo {title} {Effects of active noise
  on transition-path dynamics},\ }\href
  {https://doi.org/10.1088/2632-072X/accc69} {\bibfield  {journal} {\bibinfo
  {journal} {J. Phys. Complex.}\ }\textbf {\bibinfo {volume} {4}},\ \bibinfo
  {pages} {025005} (\bibinfo {year} {2023})}\BibitemShut {NoStop}%
\bibitem [{\citenamefont {Jauslin}(1987)}]{jauslin1987nondifferentiable}%
  \BibitemOpen
  \bibfield  {author} {\bibinfo {author} {\bibfnamefont {H.}~\bibnamefont
  {Jauslin}},\ }\bibfield  {title} {\bibinfo {title} {Nondifferentiable
  potentials for nonequilibrium steady states},\ }\href
  {https://doi.org/10.1016/0378-4371(87)90151-8} {\bibfield  {journal}
  {\bibinfo  {journal} {Physica A}\ }\textbf {\bibinfo {volume} {144}},\
  \bibinfo {pages} {179} (\bibinfo {year} {1987})}\BibitemShut {NoStop}%
\bibitem [{\citenamefont {Day}(1987)}]{day1987recent}%
  \BibitemOpen
  \bibfield  {author} {\bibinfo {author} {\bibfnamefont {M.~V.}\ \bibnamefont
  {Day}},\ }\bibfield  {title} {\bibinfo {title} {Recent progress on the small
  parameter exit problem},\ }\href {https://doi.org/10.1080/17442508708833440}
  {\bibfield  {journal} {\bibinfo  {journal} {Stochastics}\ }\textbf {\bibinfo
  {volume} {20}},\ \bibinfo {pages} {121} (\bibinfo {year} {1987})}\BibitemShut
  {NoStop}%
\bibitem [{\citenamefont {Graham}\ and\ \citenamefont
  {T{\'e}l}(1985)}]{graham1985weak}%
  \BibitemOpen
  \bibfield  {author} {\bibinfo {author} {\bibfnamefont {R.}~\bibnamefont
  {Graham}}\ and\ \bibinfo {author} {\bibfnamefont {T.}~\bibnamefont
  {T{\'e}l}},\ }\bibfield  {title} {\bibinfo {title} {Weak-noise limit of
  {Fokker-Planck} models and nondifferentiable potentials for dissipative
  dynamical systems},\ }\href {https://doi.org/10.1103/physreva.31.1109}
  {\bibfield  {journal} {\bibinfo  {journal} {Phys. Rev. A}\ }\textbf {\bibinfo
  {volume} {31}},\ \bibinfo {pages} {1109} (\bibinfo {year}
  {1985})}\BibitemShut {NoStop}%
\bibitem [{\citenamefont {Graham}\ and\ \citenamefont
  {T{\'e}l}(1986)}]{graham1986nonequilibrium}%
  \BibitemOpen
  \bibfield  {author} {\bibinfo {author} {\bibfnamefont {R.}~\bibnamefont
  {Graham}}\ and\ \bibinfo {author} {\bibfnamefont {T.}~\bibnamefont
  {T{\'e}l}},\ }\bibfield  {title} {\bibinfo {title} {Nonequilibrium potential
  for coexisting attractors},\ }\href
  {https://doi.org/10.1103/physreva.33.1322} {\bibfield  {journal} {\bibinfo
  {journal} {Phys. Rev. A}\ }\textbf {\bibinfo {volume} {33}},\ \bibinfo
  {pages} {1322} (\bibinfo {year} {1986})}\BibitemShut {NoStop}%
\bibitem [{\citenamefont {Maier}\ and\ \citenamefont
  {Stein}(1992)}]{maier1992transition}%
  \BibitemOpen
  \bibfield  {author} {\bibinfo {author} {\bibfnamefont {R.~S.}\ \bibnamefont
  {Maier}}\ and\ \bibinfo {author} {\bibfnamefont {D.~L.}\ \bibnamefont
  {Stein}},\ }\bibfield  {title} {\bibinfo {title} {{Transition-Rate Theory for
  Nongradient Drift Fields}},\ }\href
  {https://doi.org/10.1103/physrevlett.69.3691} {\bibfield  {journal} {\bibinfo
   {journal} {Phys. Rev. Lett.}\ }\textbf {\bibinfo {volume} {69}},\ \bibinfo
  {pages} {3691} (\bibinfo {year} {1992})}\BibitemShut {NoStop}%
\bibitem [{\citenamefont {Maier}\ and\ \citenamefont
  {Stein}(1993{\natexlab{a}})}]{maier1993escape}%
  \BibitemOpen
  \bibfield  {author} {\bibinfo {author} {\bibfnamefont {R.~S.}\ \bibnamefont
  {Maier}}\ and\ \bibinfo {author} {\bibfnamefont {D.~L.}\ \bibnamefont
  {Stein}},\ }\bibfield  {title} {\bibinfo {title} {Escape problem for
  irreversible systems},\ }\href {https://doi.org/10.1103/physreve.48.931}
  {\bibfield  {journal} {\bibinfo  {journal} {Phys. Rev. E}\ }\textbf {\bibinfo
  {volume} {48}},\ \bibinfo {pages} {931} (\bibinfo {year}
  {1993}{\natexlab{a}})}\BibitemShut {NoStop}%
\bibitem [{\citenamefont {Maier}\ and\ \citenamefont
  {Stein}(1993{\natexlab{b}})}]{maier1993effect}%
  \BibitemOpen
  \bibfield  {author} {\bibinfo {author} {\bibfnamefont {R.~S.}\ \bibnamefont
  {Maier}}\ and\ \bibinfo {author} {\bibfnamefont {D.~L.}\ \bibnamefont
  {Stein}},\ }\bibfield  {title} {\bibinfo {title} {{Effect of Focusing and
  Caustics on Exit Phenomena in Systems Lacking Detailed Balance}},\ }\href
  {https://doi.org/10.1103/physrevlett.71.1783} {\bibfield  {journal} {\bibinfo
   {journal} {Phys. Rev. Lett.}\ }\textbf {\bibinfo {volume} {71}},\ \bibinfo
  {pages} {1783} (\bibinfo {year} {1993}{\natexlab{b}})}\BibitemShut {NoStop}%
\bibitem [{\citenamefont {Bertini}\ \emph {et~al.}(2010)\citenamefont
  {Bertini}, \citenamefont {De~Sole}, \citenamefont {Gabrielli}, \citenamefont
  {Jona-Lasinio},\ and\ \citenamefont {Landim}}]{bertini2010lagrangian}%
  \BibitemOpen
  \bibfield  {author} {\bibinfo {author} {\bibfnamefont {L.}~\bibnamefont
  {Bertini}}, \bibinfo {author} {\bibfnamefont {A.}~\bibnamefont {De~Sole}},
  \bibinfo {author} {\bibfnamefont {D.}~\bibnamefont {Gabrielli}}, \bibinfo
  {author} {\bibfnamefont {G.}~\bibnamefont {Jona-Lasinio}},\ and\ \bibinfo
  {author} {\bibfnamefont {C.}~\bibnamefont {Landim}},\ }\bibfield  {title}
  {\bibinfo {title} {Lagrangian phase transitions in nonequilibrium
  thermodynamic systems},\ }\href
  {https://doi.org/10.1088/1742-5468/2010/11/l11001} {\bibfield  {journal}
  {\bibinfo  {journal} {J. Stat. Mech.}\ }\textbf {\bibinfo {volume} {2010}},\
  \bibinfo {pages} {L11001} (\bibinfo {year} {2010})}\BibitemShut {NoStop}%
\bibitem [{\citenamefont {Baek}\ and\ \citenamefont
  {Kafri}(2015)}]{baek2015singularities}%
  \BibitemOpen
  \bibfield  {author} {\bibinfo {author} {\bibfnamefont {Y.}~\bibnamefont
  {Baek}}\ and\ \bibinfo {author} {\bibfnamefont {Y.}~\bibnamefont {Kafri}},\
  }\bibfield  {title} {\bibinfo {title} {Singularities in large deviation
  functions},\ }\href {https://doi.org/10.1088/1742-5468/2015/08/P08026}
  {\bibfield  {journal} {\bibinfo  {journal} {J. Stat. Mech.}\ }\textbf
  {\bibinfo {volume} {2015}},\ \bibinfo {pages} {P08026} (\bibinfo {year}
  {2015})}\BibitemShut {NoStop}%
\bibitem [{\citenamefont {Nemoto}\ \emph {et~al.}(2017)\citenamefont {Nemoto},
  \citenamefont {Jack},\ and\ \citenamefont {Lecomte}}]{nemoto2017finite}%
  \BibitemOpen
  \bibfield  {author} {\bibinfo {author} {\bibfnamefont {T.}~\bibnamefont
  {Nemoto}}, \bibinfo {author} {\bibfnamefont {R.~L.}\ \bibnamefont {Jack}},\
  and\ \bibinfo {author} {\bibfnamefont {V.}~\bibnamefont {Lecomte}},\
  }\bibfield  {title} {\bibinfo {title} {{Finite-Size} {S}caling of a
  {First-Order} {D}ynamical {P}hase {T}ransition: {A}daptive {P}opulation
  {D}ynamics and an {E}ffective {M}odel},\ }\href
  {https://doi.org/10.1103/physrevlett.118.115702} {\bibfield  {journal}
  {\bibinfo  {journal} {Phys. Rev. Lett.}\ }\textbf {\bibinfo {volume} {118}},\
  \bibinfo {pages} {115702} (\bibinfo {year} {2017})}\BibitemShut {NoStop}%
\bibitem [{\citenamefont {Baek}\ \emph {et~al.}(2018)\citenamefont {Baek},
  \citenamefont {Kafri},\ and\ \citenamefont {Lecomte}}]{baek2018dynamical}%
  \BibitemOpen
  \bibfield  {author} {\bibinfo {author} {\bibfnamefont {Y.}~\bibnamefont
  {Baek}}, \bibinfo {author} {\bibfnamefont {Y.}~\bibnamefont {Kafri}},\ and\
  \bibinfo {author} {\bibfnamefont {V.}~\bibnamefont {Lecomte}},\ }\bibfield
  {title} {\bibinfo {title} {Dynamical phase transitions in the current
  distribution of driven diffusive channels},\ }\href
  {https://doi.org/10.1088/1751-8121/aaa8f9} {\bibfield  {journal} {\bibinfo
  {journal} {J. Phys. A: Math. Theor.}\ }\textbf {\bibinfo {volume} {51}},\
  \bibinfo {pages} {105001} (\bibinfo {year} {2018})}\BibitemShut {NoStop}%
\bibitem [{\citenamefont {Aminov}\ \emph {et~al.}(2014)\citenamefont {Aminov},
  \citenamefont {Bunin},\ and\ \citenamefont
  {Kafri}}]{aminov2014singularities}%
  \BibitemOpen
  \bibfield  {author} {\bibinfo {author} {\bibfnamefont {A.}~\bibnamefont
  {Aminov}}, \bibinfo {author} {\bibfnamefont {G.}~\bibnamefont {Bunin}},\ and\
  \bibinfo {author} {\bibfnamefont {Y.}~\bibnamefont {Kafri}},\ }\bibfield
  {title} {\bibinfo {title} {Singularities in large deviation functionals of
  bulk-driven transport models},\ }\href
  {https://doi.org/10.1088/1742-5468/2014/08/p08017} {\bibfield  {journal}
  {\bibinfo  {journal} {J. Stat. Mech.}\ }\textbf {\bibinfo {volume} {2014}},\
  \bibinfo {pages} {P08017} (\bibinfo {year} {2014})}\BibitemShut {NoStop}%
\bibitem [{\citenamefont {Bouchet}\ \emph {et~al.}(2016)\citenamefont
  {Bouchet}, \citenamefont {Gaw\k{e}dzki},\ and\ \citenamefont
  {Nardini}}]{bouchet2016perturbative}%
  \BibitemOpen
  \bibfield  {author} {\bibinfo {author} {\bibfnamefont {F.}~\bibnamefont
  {Bouchet}}, \bibinfo {author} {\bibfnamefont {K.}~\bibnamefont
  {Gaw\k{e}dzki}},\ and\ \bibinfo {author} {\bibfnamefont {C.}~\bibnamefont
  {Nardini}},\ }\bibfield  {title} {\bibinfo {title} {Perturbative
  {C}alculation of {Quasi-Potential} in {N}on-equilibrium {D}iffusions: A
  {Mean-Field} {E}xample},\ }\href {https://doi.org/10.1007/s10955-016-1503-2}
  {\bibfield  {journal} {\bibinfo  {journal} {J. Stat. Phys.}\ }\textbf
  {\bibinfo {volume} {163}},\ \bibinfo {pages} {1157} (\bibinfo {year}
  {2016})}\BibitemShut {NoStop}%
\bibitem [{\citenamefont {Grafke}\ and\ \citenamefont
  {Vanden-Eijnden}(2019)}]{grafke2019numerical}%
  \BibitemOpen
  \bibfield  {author} {\bibinfo {author} {\bibfnamefont {T.}~\bibnamefont
  {Grafke}}\ and\ \bibinfo {author} {\bibfnamefont {E.}~\bibnamefont
  {Vanden-Eijnden}},\ }\bibfield  {title} {\bibinfo {title} {Numerical
  computation of rare events via large deviation theory},\ }\href
  {https://doi.org/10.1063/1.5084025} {\bibfield  {journal} {\bibinfo
  {journal} {Chaos}\ }\textbf {\bibinfo {volume} {29}},\ \bibinfo {pages}
  {063118} (\bibinfo {year} {2019})}\BibitemShut {NoStop}%
\bibitem [{\citenamefont {Kikuchi}\ \emph {et~al.}(2020)\citenamefont
  {Kikuchi}, \citenamefont {Singh}, \citenamefont {Cates},\ and\ \citenamefont
  {Adhikari}}]{kikuchi2020ritz}%
  \BibitemOpen
  \bibfield  {author} {\bibinfo {author} {\bibfnamefont {L.}~\bibnamefont
  {Kikuchi}}, \bibinfo {author} {\bibfnamefont {R.}~\bibnamefont {Singh}},
  \bibinfo {author} {\bibfnamefont {M.~E.}\ \bibnamefont {Cates}},\ and\
  \bibinfo {author} {\bibfnamefont {R.}~\bibnamefont {Adhikari}},\ }\bibfield
  {title} {\bibinfo {title} {Ritz method for transition paths and
  quasipotentials of rare diffusive events},\ }\href
  {https://doi.org/10.1103/physrevresearch.2.033208} {\bibfield  {journal}
  {\bibinfo  {journal} {Phys. Rev. Research}\ }\textbf {\bibinfo {volume}
  {2}},\ \bibinfo {pages} {033208} (\bibinfo {year} {2020})}\BibitemShut
  {NoStop}%
\bibitem [{\citenamefont {Bouchet}\ and\ \citenamefont
  {Reygner}(2016)}]{bouchet2016generalisation}%
  \BibitemOpen
  \bibfield  {author} {\bibinfo {author} {\bibfnamefont {F.}~\bibnamefont
  {Bouchet}}\ and\ \bibinfo {author} {\bibfnamefont {J.}~\bibnamefont
  {Reygner}},\ }\bibfield  {title} {\bibinfo {title} {Generalisation of the
  {Eyring}--{Kramers} {T}ransition {R}ate {F}ormula to {I}rreversible
  {D}iffusion {P}rocesses},\ }\href {https://doi.org/10.1007/s00023-016-0507-4}
  {\bibfield  {journal} {\bibinfo  {journal} {Ann. Henri Poincaré}\ }\textbf
  {\bibinfo {volume} {17}},\ \bibinfo {pages} {3499} (\bibinfo {year}
  {2016})}\BibitemShut {NoStop}%
\bibitem [{\citenamefont {Maier}\ and\ \citenamefont
  {Stein}(1996)}]{maier1996scaling}%
  \BibitemOpen
  \bibfield  {author} {\bibinfo {author} {\bibfnamefont {R.~S.}\ \bibnamefont
  {Maier}}\ and\ \bibinfo {author} {\bibfnamefont {D.~L.}\ \bibnamefont
  {Stein}},\ }\bibfield  {title} {\bibinfo {title} {A scaling theory of
  bifurcations in the symmetric weak-noise escape problem},\ }\href
  {https://doi.org/10.1007/bf02183736} {\bibfield  {journal} {\bibinfo
  {journal} {J. Stat. Phys.}\ }\textbf {\bibinfo {volume} {83}},\ \bibinfo
  {pages} {291} (\bibinfo {year} {1996})}\BibitemShut {NoStop}%
\bibitem [{\citenamefont {Feng}\ and\ \citenamefont
  {Kurtz}(2006)}]{feng2006large}%
  \BibitemOpen
  \bibfield  {author} {\bibinfo {author} {\bibfnamefont {J.}~\bibnamefont
  {Feng}}\ and\ \bibinfo {author} {\bibfnamefont {T.}~\bibnamefont {Kurtz}},\
  }\href {https://doi.org/10.1090/surv/131} {\emph {\bibinfo {title} {{Large
  Deviations for Stochastic Processes}}}},\ Mathematical Surveys and
  Monographs\ (\bibinfo  {publisher} {American Mathematical Society},\ \bibinfo
  {year} {2006})\BibitemShut {NoStop}%
\bibitem [{\citenamefont {Rey-Bellet}\ and\ \citenamefont
  {Spiliopoulos}(2016)}]{spilio2016}%
  \BibitemOpen
  \bibfield  {author} {\bibinfo {author} {\bibfnamefont {L.}~\bibnamefont
  {Rey-Bellet}}\ and\ \bibinfo {author} {\bibfnamefont {K.}~\bibnamefont
  {Spiliopoulos}},\ }\bibfield  {title} {\bibinfo {title} {Improving the
  {C}onvergence of {R}eversible {S}amplers},\ }\href
  {https://doi.org/10.1007/s10955-016-1565-1} {\bibfield  {journal} {\bibinfo
  {journal} {J. Stat. Phys.}\ }\textbf {\bibinfo {volume} {164}},\ \bibinfo
  {pages} {472} (\bibinfo {year} {2016})}\BibitemShut {NoStop}%
\bibitem [{\citenamefont {Kaiser}\ \emph {et~al.}(2017)\citenamefont {Kaiser},
  \citenamefont {Jack},\ and\ \citenamefont {Zimmer}}]{kaiser2017acceleration}%
  \BibitemOpen
  \bibfield  {author} {\bibinfo {author} {\bibfnamefont {M.}~\bibnamefont
  {Kaiser}}, \bibinfo {author} {\bibfnamefont {R.~L.}\ \bibnamefont {Jack}},\
  and\ \bibinfo {author} {\bibfnamefont {J.}~\bibnamefont {Zimmer}},\
  }\bibfield  {title} {\bibinfo {title} {Acceleration of {C}onvergence to
  {E}quilibrium in {M}arkov {C}hains by {B}reaking {D}etailed {B}alance},\
  }\href {https://doi.org/10.1007/s10955-017-1805-z} {\bibfield  {journal}
  {\bibinfo  {journal} {J. Stat. Phys.}\ }\textbf {\bibinfo {volume} {168}},\
  \bibinfo {pages} {259} (\bibinfo {year} {2017})}\BibitemShut {NoStop}%
\bibitem [{\citenamefont {B\"orner}\ \emph {et~al.}(2024)\citenamefont
  {B\"orner}, \citenamefont {Deeley}, \citenamefont {R\"omer}, \citenamefont
  {Grafke}, \citenamefont {Lucarini},\ and\ \citenamefont
  {Feudel}}]{borner2024saddle}%
  \BibitemOpen
  \bibfield  {author} {\bibinfo {author} {\bibfnamefont {R.}~\bibnamefont
  {B\"orner}}, \bibinfo {author} {\bibfnamefont {R.}~\bibnamefont {Deeley}},
  \bibinfo {author} {\bibfnamefont {R.}~\bibnamefont {R\"omer}}, \bibinfo
  {author} {\bibfnamefont {T.}~\bibnamefont {Grafke}}, \bibinfo {author}
  {\bibfnamefont {V.}~\bibnamefont {Lucarini}},\ and\ \bibinfo {author}
  {\bibfnamefont {U.}~\bibnamefont {Feudel}},\ }\bibfield  {title} {\bibinfo
  {title} {Saddle avoidance of noise-induced transitions in multiscale
  systems},\ }\href {https://doi.org/10.1103/PhysRevResearch.6.L042053}
  {\bibfield  {journal} {\bibinfo  {journal} {Phys. Rev. Res.}\ }\textbf
  {\bibinfo {volume} {6}},\ \bibinfo {pages} {L042053} (\bibinfo {year}
  {2024})}\BibitemShut {NoStop}%
\bibitem [{\citenamefont {E}\ \emph {et~al.}(2004)\citenamefont {E},
  \citenamefont {Ren},\ and\ \citenamefont
  {Vanden‐Eijnden}}]{weinan2004minimum}%
  \BibitemOpen
  \bibfield  {author} {\bibinfo {author} {\bibfnamefont {W.}~\bibnamefont {E}},
  \bibinfo {author} {\bibfnamefont {W.}~\bibnamefont {Ren}},\ and\ \bibinfo
  {author} {\bibfnamefont {E.}~\bibnamefont {Vanden‐Eijnden}},\ }\bibfield
  {title} {\bibinfo {title} {Minimum action method for the study of rare
  events},\ }\href {https://doi.org/10.1002/cpa.20005} {\bibfield  {journal}
  {\bibinfo  {journal} {Commun. Pure Appl. Math.}\ }\textbf {\bibinfo {volume}
  {57}},\ \bibinfo {pages} {637} (\bibinfo {year} {2004})}\BibitemShut
  {NoStop}%
\bibitem [{\citenamefont {Heymann}\ and\ \citenamefont
  {Vanden‐Eijnden}(2008)}]{heymann2008geometric}%
  \BibitemOpen
  \bibfield  {author} {\bibinfo {author} {\bibfnamefont {M.}~\bibnamefont
  {Heymann}}\ and\ \bibinfo {author} {\bibfnamefont {E.}~\bibnamefont
  {Vanden‐Eijnden}},\ }\bibfield  {title} {\bibinfo {title} {The geometric
  minimum action method: {A} least action principle on the space of curves},\
  }\href {https://doi.org/10.1002/cpa.20238} {\bibfield  {journal} {\bibinfo
  {journal} {Commun. Pure Appl. Math.}\ }\textbf {\bibinfo {volume} {61}},\
  \bibinfo {pages} {1052} (\bibinfo {year} {2008})}\BibitemShut {NoStop}%
\bibitem [{\citenamefont {Grafke}\ \emph
  {et~al.}(2017{\natexlab{b}})\citenamefont {Grafke}, \citenamefont
  {Sch{\"a}fer},\ and\ \citenamefont {Vanden-Eijnden}}]{grafke2017long}%
  \BibitemOpen
  \bibfield  {author} {\bibinfo {author} {\bibfnamefont {T.}~\bibnamefont
  {Grafke}}, \bibinfo {author} {\bibfnamefont {T.}~\bibnamefont
  {Sch{\"a}fer}},\ and\ \bibinfo {author} {\bibfnamefont {E.}~\bibnamefont
  {Vanden-Eijnden}},\ }\bibfield  {title} {\bibinfo {title} {{Long Term Effects
  of Small Random Perturbations on Dynamical Systems: Theoretical and
  Computational Tools}},\ }in\ \href
  {https://doi.org/10.1007/978-1-4939-6969-2_2} {\emph {\bibinfo {booktitle}
  {Recent Progress and Modern Challenges in Applied Mathematics, Modeling and
  Computational Science}}},\ \bibinfo {editor} {edited by\ \bibinfo {editor}
  {\bibfnamefont {R.}~\bibnamefont {Melnik}}}\ (\bibinfo  {publisher}
  {Springer},\ \bibinfo {address} {New York},\ \bibinfo {year} {2017})\ pp.\
  \bibinfo {pages} {17--55}\BibitemShut {NoStop}%
\bibitem [{\citenamefont {Heymann}\ and\ \citenamefont
  {Vanden-Eijnden}(2008)}]{heymann2008pathways}%
  \BibitemOpen
  \bibfield  {author} {\bibinfo {author} {\bibfnamefont {M.}~\bibnamefont
  {Heymann}}\ and\ \bibinfo {author} {\bibfnamefont {E.}~\bibnamefont
  {Vanden-Eijnden}},\ }\bibfield  {title} {\bibinfo {title} {{Pathways of
  Maximum Likelihood for Rare Events in Nonequilibrium Systems: Application to
  Nucleation in the Presence of Shear}},\ }\href
  {https://doi.org/10.1103/physrevlett.100.140601} {\bibfield  {journal}
  {\bibinfo  {journal} {Phys. Rev. Lett.}\ }\textbf {\bibinfo {volume} {100}},\
  \bibinfo {pages} {140601} (\bibinfo {year} {2008})}\BibitemShut {NoStop}%
\bibitem [{\citenamefont {Zakine}\ and\ \citenamefont
  {Vanden-Eijnden}(2023)}]{zakine2023minimum}%
  \BibitemOpen
  \bibfield  {author} {\bibinfo {author} {\bibfnamefont {R.}~\bibnamefont
  {Zakine}}\ and\ \bibinfo {author} {\bibfnamefont {E.}~\bibnamefont
  {Vanden-Eijnden}},\ }\bibfield  {title} {\bibinfo {title} {{Minimum-Action
  Method for Nonequilibrium Phase Transitions}},\ }\href
  {https://doi.org/10.1103/physrevx.13.041044} {\bibfield  {journal} {\bibinfo
  {journal} {Phys. Rev. X}\ }\textbf {\bibinfo {volume} {13}},\ \bibinfo
  {pages} {041044} (\bibinfo {year} {2023})}\BibitemShut {NoStop}%
\end{thebibliography}%

\end{document}